\documentclass[a4paper,fleqn]{cas-sc}

\usepackage[T1]{fontenc}
\usepackage{amsmath,amssymb}
\usepackage[numbers,sort&compress]{natbib}
\usepackage{graphicx}
\usepackage{booktabs}
\usepackage{tabularx}
\usepackage{placeins}
\usepackage{float}
\ExplSyntaxOn
\cs_set:Npn \__first_footerline:
  { \group_begin: \small \sffamily \__short_authors: \group_end: }
\ExplSyntaxOff

\providecommand{\doi}[1]{%
  \href{https://doi.org/#1}{\nolinkurl{#1}}%
}

\begin{document}

\let\WriteBookmarks\relax
\def\floatpagepagefraction{1}
\def\textpagefraction{.001}

\shorttitle{Pauli-consistent ensemble Monte Carlo method for bipolar charge transport in graphene}
\shortauthors{T. Zalinyan and G. Nastasi}

\title[mode=title]{Pauli-consistent ensemble Monte Carlo method for bipolar charge transport in graphene}

\author[1]{Tigran Zalinyan}[orcid=0009-0008-2304-1637]
\cormark[1]
\ead{tigran.zalinyan@ysu.am}

\author[2]{Giovanni Nastasi}[orcid=0000-0002-6374-8441]
\ead{giovanni.nastasi@unikore.it}

\affiliation[1]{organization={Research Institute of Physics, Yerevan State University},
  city={Yerevan}, country={Armenia}}
\affiliation[2]{organization={Department of Engineering and Architecture, University of Enna ``Kore''},
  city={Enna}, country={Italy}}

\cortext[1]{Corresponding author}

\begin{abstract}
We develop a Pauli-consistent ensemble Monte Carlo method for
bipolar charge transport in graphene, with an explicit treatment
of phonon-assisted interband scattering, including electron--hole
generation and recombination.
The method employs a valence-vacancy representation to describe
coupled electron and hole populations without explicitly simulating
the highly populated valence band.
It preserves the discrete occupation bounds during both free
flight and scattering while conserving net charge.
The numerical properties of the method are assessed through
sensitivity and refinement studies and comparison with an independent
deterministic discontinuous Galerkin solution of the same kinetic model.
The two approaches show very good agreement for the time evolution,
carrier distributions, and steady-state transport observables across
the investigated transport regimes.
The results also demonstrate the computational practicality of
the method for microscopic simulations of bipolar graphene transport.
\end{abstract}


\maketitle

\section{Introduction}
\label{sec:introduction}

Monte Carlo methods are widely used for the numerical study
of carrier transport in semiconductors
\cite{JacoboniLugli1989MonteCarloMethod,Jacoboni2010TheoryOfElectronTransport}. One of their main limitations
is statistical noise, whose reduction generally requires
larger particle ensembles and consequently increases the
computational cost. Modern high-performance computing
resources nevertheless make simulations with large
particle ensembles feasible.

For degenerate carrier systems, a further numerical
difficulty is associated with the Pauli exclusion
principle. The usual rejection technique accounts
for the occupation of the proposed final state during
a scattering event \cite{LugliFerry1985}, but this
condition alone does not prevent the reconstructed
distribution from exceeding the admissible bound
during particle free flight
\cite{Tadyszak1996,BorowikThobel1998}.
This difficulty was addressed by a Pauli-consistent
direct simulation Monte Carlo (DSMC) scheme
\cite{Romano2015}, subsequently referred
to as the ``new'' ensemble Monte Carlo (NEMC) method.
The original formulation considered unipolar
conduction-band transport with intraband
electron--phonon scattering.
By translating the occupation grid together with
the particle ensemble during drift and applying
final-state Pauli tests during scattering, NEMC
preserves the discrete occupation bound during both substeps.
Further developments of the Pauli treatment and
numerical implementation of NEMC were presented
in Ref.~\cite{Coco2021}.
Electron--electron scattering was subsequently
incorporated into NEMC, with both intraband and
interband processes examined, although numerical
simulations were restricted to the unipolar case
\cite{NastasiRomano2023}.
The NEMC treatment of intraband electron--electron
scattering was further developed in
Ref.~\cite{ZalinyanNastasi2026}, where a sampled-partner
approximation was introduced to reduce the computational
cost of scattering-rate evaluation while retaining
the event-level collision treatment, thereby enabling
large-ensemble, low-noise simulations.
Related particle-based approaches for uncertainty
quantification of graphene-nanoribbon transport
have also been developed
\cite{Medaglia2025UncertaintyQuantificationForCharge}.

Near charge neutrality in graphene, electrons and holes
can contribute simultaneously to charge transport.
Interband scattering couples the two carrier populations
through electron--hole generation and recombination.
Phonon-assisted transitions provide one mechanism for
this coupling \cite{Rana2009Recombination}.
A bipolar solution of the semiclassical Boltzmann
equations has shown that neglecting interband scattering
can substantially affect the predicted transport
properties, with its influence depending on the
Fermi energy and the transport regime
\cite{MajoranaNastasiRomano2019}.
Related deterministic Boltzmann studies address
charge transport in graphene nanoribbons
\cite{NastasiCamiolaRomano2022GrapheneNanoribbons}
and the time-dependent collision operator for
two-dimensional materials
\cite{Wadgaonkar2022NumericalSolverForThe}.

Bipolar and interband transport have also been studied
within other Monte Carlo frameworks. Device simulations
have included band-to-band tunnelling
\cite{David2012Semiclassical,Paussa2014BTBT}
and the simultaneous transport of electrons and holes
supplied by contacts \cite{Truccolo2026Scattering}.
Earlier ensemble Monte Carlo studies have also examined
carrier transport and photoexcited-carrier dynamics
in graphene
\cite{Rengel2013DiffusionCoefficientCorrelationFunction,
Iglesias2015CarrierCarrierAndCarrier,
Iglesias2016HotCarrierAndHot}.
Scattering-induced ambipolar transport, including
phonon-assisted generation and recombination as well
as Coulomb-induced interband processes, has also
been investigated using a conventional ensemble
Monte Carlo formulation \cite{Iglesias2019}.
However, in the models used so far, the Pauli exclusion
principle is generally imposed through standard final-state
rejection during scattering, which does not guarantee the
occupation bound during free flight.
Further numerical challenges, particularly in the explicit
treatment of interband generation and recombination,
provide additional motivation for the present work.

In extending NEMC to bipolar transport, the representation
of the valence band is a central issue. Explicitly
simulating its highly populated electronic states can
require a prohibitively large particle ensemble
\cite{MajoranaNastasiRomano2019}. Representing the valence
band through holes reduces this burden, but generation must
then account for occupied valence states that are not
explicitly simulated, while recombination requires the
consistent removal of an electron--hole pair. These
population changes must be incorporated into the numerical
evolution while preserving the admissible occupations of
both bands.

The aim of the present work is to formulate and systematically
assess a Pauli-consistent bipolar NEMC framework for charge
transport in graphene, including phonon-assisted interband
generation and recombination. We consider spatially homogeneous
transport in suspended monolayer graphene under a constant
electric field. The numerical properties of the method are
investigated through numerical-sensitivity and refinement
studies and cross-comparison with an independent deterministic
discontinuous Galerkin (DG) solution of the same kinetic model,
with particular attention to numerical agreement and
computational cost.

The paper is organized as follows. Section~\ref{sec:methods}
introduces the physical model and the bipolar NEMC method.
Section~\ref{sec:results} presents the numerical results
and comparisons with DG. Section~\ref{sec:discussion}
discusses the results and limitations of the approach,
while Section~\ref{sec:conclusions} contains the conclusions.
Additional details and supporting results are given
in the appendices.

\section{Methods}\label{sec:methods}

\subsection{Physical model and kinetic equations}
\label{sec:physical}

We consider spatially homogeneous charge transport in
suspended monolayer graphene under a constant electric
field. Electrons in the conduction and valence bands
are described by the semiclassical Boltzmann equations.
For simplicity, and to focus on the development of the
bipolar NEMC method, we restrict the collision model
to electron--phonon scattering. The phonon bath is
assumed to remain at thermal equilibrium at the lattice
temperature $T$, and both intraband and kinematically
allowed phonon-assisted interband transitions are included.
Carrier--carrier scattering, Landau--Zener interband
tunnelling, nonequilibrium phonon dynamics, and lattice
heating are not included in the present study.

In the low-energy Dirac approximation, graphene is
described by gapless conduction and valence bands
with linear energy dispersion near the two inequivalent
valleys
\cite{DasSarma2011ElectronicTransportInTwo,
Katsnelson2020PhysicsOfGraphene,CastroNeto2009ElectronicPropertiesOfGraphene}.
These valleys are assumed to have equal occupations.
The wave vector $\mathbf{k}$ is measured from the
center of the corresponding valley. Although the
first Brillouin zone is hexagonal \cite{Grundmann2021PhysicsOfSemiconductors}, in the continuum
approximation we take $\mathbf{k}\in\mathbb{R}^2$.
The Fermi energy determines the equilibrium
occupations of the two bands and the relative
electron and hole populations.
The band energy and group velocity are
\begin{equation}
   \varepsilon_s(\mathbf{k})=s\hbar v_F|\mathbf{k}|,
   \qquad
   \mathbf{v}_s(\mathbf{k})
   =\frac{1}{\hbar}\nabla_{\mathbf{k}}\varepsilon_s(\mathbf{k})
   =s v_F\frac{\mathbf{k}}{|\mathbf{k}|},
   \qquad s=\pm1,
   \label{eq:band-dispersion}
\end{equation}
where $s=+1$ and $s=-1$ denote the conduction and
valence bands, respectively, $v_F$ is the Fermi velocity,
and $\hbar$ is the reduced Planck constant.
The velocity expression applies for $\mathbf{k}\ne\mathbf{0}$.
We denote by $f_s(t,\mathbf{k})$ the dimensionless
electron occupation of band $s$, with $0\leq f_s\leq1$.

The evolution of the two distributions is governed by
\begin{equation}
   \frac{\partial f_s}{\partial t}
   -\frac{e}{\hbar}\mathbf{E}\cdot\nabla_{\mathbf{k}}f_s
   =Q_s[f_+,f_-],
   \qquad s=\pm1,
   \label{eq:bipolar-bte}
\end{equation}
where $e>0$ is the elementary charge and
$\mathbf{E}\in\mathbb{R}^2$ is the applied electric field.
The electron--phonon collision operator has the gain--loss form
\begin{equation}
Q_s[f_+,f_-](\mathbf{k})
=
\sum_{s'=\pm1}
\int_{\mathbb R^2}
\left\{
S_{s's}(\mathbf{k}',\mathbf{k})
f_{s'}(\mathbf{k}')
[1-f_s(\mathbf{k})]
-
S_{ss'}(\mathbf{k},\mathbf{k}')
f_s(\mathbf{k})
[1-f_{s'}(\mathbf{k}')]
\right\}
\,\mathrm{d}\mathbf{k}'.
\label{eq:eph-collision-operator}
\end{equation}
Here the time argument has been omitted for brevity, and $S_{ss'}(\mathbf{k},\mathbf{k}')$ denotes the transition kernel from $(s,\mathbf{k})$ to $(s',\mathbf{k}')$. The factors $1-f_s(\mathbf{k})$ and $1-f_{s'}(\mathbf{k}')$ account for final-state availability according to the Pauli exclusion principle. Terms with $s'=s$ describe intraband scattering, whereas terms with $s'\ne s$ couple the two bands.

The electron--phonon model contains an effective intravalley acoustic channel, an effective intravalley optical channel combining the longitudinal optical (LO) and transverse optical (TO) contributions, and an intervalley $K$ channel. Acoustic scattering is treated as elastic. The optical and intervalley modes are taken to be dispersionless, with phonon energies $\hbar\omega_\nu$ and equilibrium Bose occupations
\begin{equation}
   n_\nu=
   \frac{1}{\exp\!\left(\hbar\omega_\nu/(k_BT)\right)-1},
   \qquad \nu\in\{O,K\},
   \label{eq:phonon-occupation}
\end{equation}
where $k_B$ is the Boltzmann constant.

The inelastic channels include both intraband and interband
transitions. In an interband absorption event, a valence
electron is promoted to the conduction band and an
electron--hole pair is generated. The reverse transition
occurs through phonon emission and corresponds to pair
recombination
\cite{Rana2009Recombination}.
Energy conservation requires
\begin{equation}
   \varepsilon_+(\mathbf{k}_c)-\varepsilon_-(\mathbf{k}_v)
   =\hbar\omega_\nu,
   \label{eq:interband-energy-conservation}
\end{equation}
where $\mathbf{k}_c$ and $\mathbf{k}_v$ denote the conduction-
and valence-band states involved. With the linear dispersion,
this condition becomes
$\hbar v_F(|\mathbf{k}_c|+|\mathbf{k}_v|)
=\hbar\omega_\nu$, restricting interband transitions
to states satisfying this energy relation.
The explicit transition kernels, scattering rates,
material parameters, and their source references are
given in Appendix~\ref{app:phonons}.

The initial distributions are taken to be Fermi--Dirac distributions at the lattice temperature,
\begin{equation}
   f_s(0,\mathbf{k})=
   \frac{1}{
   1+\exp\!\left[
   \frac{\varepsilon_s(\mathbf{k})-\varepsilon_F}{k_BT}
   \right]},
   \qquad s=\pm1,
   \label{eq:initial-fermi-dirac}
\end{equation}
where $\varepsilon_F$ is the initial Fermi energy measured from the Dirac point.

For the bipolar formulation, it is useful to distinguish explicitly between the valence-vacancy coordinate and the physical hole coordinate. We define
\begin{equation}
   F(t,\mathbf{k})=f_+(t,\mathbf{k}),
   \qquad
   G(t,\boldsymbol{\kappa})=1-f_-(t,\boldsymbol{\kappa}),
   \label{eq:FG-definitions}
\end{equation}
where $F$ is the conduction-electron occupation and $G$ is the valence-vacancy occupation expressed in the valence-electron wave-vector coordinate $\boldsymbol{\kappa}$. A physical hole associated with a vacancy at $\boldsymbol{\kappa}$ has wave vector $\mathbf{k}_h=-\boldsymbol{\kappa}$. Accordingly, the physical electron and hole distributions are
\begin{equation}
   f^e(t,\mathbf{k})=F(t,\mathbf{k}),
   \qquad
   f^h(t,\mathbf{k})=G(t,-\mathbf{k})
   =1-f_-(t,-\mathbf{k}).
   \label{eq:physical-carrier-distributions}
\end{equation}
In these physical electron--hole coordinates, both carrier species have the positive energy $\varepsilon(\mathbf{k})=\hbar v_F|\mathbf{k}|$ and velocity $\mathbf{v}(\mathbf{k})=v_F\mathbf{k}/|\mathbf{k}|$.

The electron and hole densities are
\begin{equation}
   n_\alpha(t)=\frac{g_sg_v}{(2\pi)^2}
   \int_{\mathbb R^2}
   f^\alpha(t,\mathbf{k})\,\mathrm{d}\mathbf{k},
   \qquad
   \alpha\in\{e,h\},
   \label{eq:carrier-observables}
\end{equation}
where $g_s=g_v=2$ are the spin and valley degeneracies. The corresponding mean velocity and mean energy are
\begin{equation}
\overline{\mathbf v}_\alpha(t)
=
\frac{g_sg_v}{(2\pi)^2n_\alpha(t)}
\int_{\mathbb R^2}
\mathbf v(\mathbf{k})f^\alpha(t,\mathbf{k})\,\mathrm{d}\mathbf{k},
\qquad
\overline{\varepsilon}_\alpha(t)
=
\frac{g_sg_v}{(2\pi)^2n_\alpha(t)}
\int_{\mathbb R^2}
\varepsilon(\mathbf{k})f^\alpha(t,\mathbf{k})\,\mathrm{d}\mathbf{k}.
\label{eq:carrier-mean-observables}
\end{equation}
The net charge density, measured relative to the filled valence band, and the total current density are
\begin{equation}
   \rho=e(n_h-n_e),
   \qquad
   \mathbf{J}
   =-e n_e\overline{\mathbf{v}}_e
    +e n_h\overline{\mathbf{v}}_h.
   \label{eq:charge-current}
\end{equation}

\subsection{The Pauli-consistent ensemble Monte Carlo method}
\label{sec:nemc}

Transport is simulated using the Pauli-consistent
``new'' ensemble Monte Carlo (NEMC) method
\cite{Romano2015,Coco2021,ZalinyanNastasi2026}.
We first recall the drift--collision construction and
the occupation normalization for a single electron
population, and then extend the method to coupled
electron and hole populations.

The initial distributions, electric-field drift, and
scattering processes determine the wave-vector region
relevant to the simulation. We truncate wave-vector
space to a square domain $\Omega_k$ and partition it
into $N_k\times N_k$ uniform square cells $\mathcal C_a$
of side $\Delta k$ and area $(\Delta k)^2$, where a
single index $a$ labels the two-dimensional cells.
The domain is chosen sufficiently large to contain
the relevant carrier distributions and valence states
contributing to interband generation. Its numerical
sensitivity is examined in
Table~\ref{tab:d-checks}.
The simulated particles have continuous wave-vector
coordinates; the grid is used to define the cellwise
occupation entering the Pauli treatment.
All simulated particles have the same statistical
weight. If $\mathrm{occ}_a(t)$ denotes the number
of particles assigned to cell $\mathcal C_a$,
the corresponding occupation fraction is
\begin{equation}
   f_a(t)=\frac{\mathrm{occ}_a(t)}{M},
   \qquad
   w_p=\frac{g_sg_v(\Delta k)^2}{(2\pi)^2M},
   \label{eq:nemc-occupation}
\end{equation}
where $M$ is the fixed maximum number of simulated
particles allowed in a cell and $w_p$ is the carrier-density
weight represented by one simulated particle.
The carrier density is $n=w_pN_p$, with
$N_p=\sum_a\mathrm{occ}_a$, consistent with
Eq.~\eqref{eq:carrier-observables}.
The quantities $M$ and $w_p$ remain fixed throughout
the simulation, and the Pauli constraint is equivalent
to $0\leq\mathrm{occ}_a\leq M$ for every cell.

We introduce a uniform time grid $t_r=r\Delta t$ and use a first-order drift--collision splitting. Over each interval $[t_r,t_{r+1}]$, the drift subproblem
\begin{equation}
   \frac{\partial f}{\partial t}
   -\frac{e}{\hbar}\mathbf{E}\cdot\nabla_{\mathbf{k}}f=0
   \label{eq:nemc-drift}
\end{equation}
is first advanced over $\Delta t$. The resulting distribution is then used as the initial condition for the collision subproblem
\begin{equation}
   \frac{\partial f}{\partial t}
   =Q_{\mathrm{intra}}[f],
   \label{eq:nemc-collision}
\end{equation}
which is advanced over the same interval $\Delta t$. Here $Q_{\mathrm{intra}}$ denotes the intraband electron--phonon collision operator for the single-population construction considered in this part.

For a constant electric field, the drift equation is solved exactly by a rigid translation in wave-vector space. The superscript $\star$ denotes the state after the drift substep. Over one time step, every particle undergoes the same displacement,
\begin{equation}
   \mathbf{k}_i^\star
   =
   \mathbf{k}_i^r
   -\frac{e}{\hbar}\mathbf{E}\Delta t.
   \label{eq:nemc-particle-drift}
\end{equation}
The occupation cells undergo the same displacement. To avoid using an increasingly displaced computational domain, NEMC represents this common translation by a Lagrangian moving grid adapted to the displaced cell positions,
\begin{equation}
   \mathcal{C}_a^\star
   =
   \mathcal{C}_a^r
   -\frac{e}{\hbar}\mathbf{E}\Delta t.
   \label{eq:nemc-grid-translation}
\end{equation}
Because the particles and the occupation grid undergo the same translation, their relative positions are unchanged and
\begin{equation}
   \mathrm{occ}_a^\star=\mathrm{occ}_a^r .
   \label{eq:nemc-drift-occupation}
\end{equation}
The cell occupations and Pauli bounds are preserved during drift without redistribution of particles among the occupation cells.

During the collision substep, the occupation grid is
kept fixed at its post-drift position. A scattering
mechanism and a candidate final wave vector are
generated from the electron--phonon transition rates
given in Appendix~\ref{app:phonons}, before applying
the final-state Pauli factor.
If the incident particle belongs to cell $\mathcal C_a$ and the proposed final state lies in cell $\mathcal C_b$, the effective occupation of the destination cell and the corresponding acceptance probability are
\begin{equation}
   \mathrm{occ}_b^{\mathrm{eff}}
   =
   \mathrm{occ}_b-\delta_{ab},
   \qquad
   P_{\mathrm{acc}}
   =
   1-\frac{\mathrm{occ}_b^{\mathrm{eff}}}{M}.
   \label{eq:nemc-intraband-acceptance}
\end{equation}
Here $\delta_{ab}$ is the Kronecker delta. When the source and destination cells coincide, the subtraction excludes the incident particle itself from the occupation used in the final-state Pauli test~\cite{ZalinyanNastasi2026}. The proposed transition is accepted with probability $P_{\mathrm{acc}}$.

If the proposed transition is accepted, the particle wave vector is replaced by the sampled final wave vector and the affected cell counts are updated immediately. For a transition between different cells, one particle is removed from the source cell and added to the destination cell. If the initial and final states belong to the same cell, the particle wave vector changes but the cell occupation remains unchanged. If the transition is rejected by the Pauli test, the particle state and all cell occupations remain unchanged.

Starting from an admissible occupation field, both substeps preserve the cell-capacity constraint and therefore the Pauli bound $0\leq f_a\leq1$. The bipolar extension retains the same drift--collision construction and introduces coupled electron and hole populations together with pair generation and recombination.

\subsection{Bipolar NEMC with interband electron--phonon scattering}
\label{sec:bipolar}

Interband scattering couples the electron and hole populations. A transition from the valence band to the conduction band creates an electron--hole pair, whereas the reverse transition removes a pair. In the NEMC representation, holes are stored in the valence-electron coordinate $\boldsymbol{\kappa}$ introduced in Section~\ref{sec:physical}, with physical hole wave vector $\mathbf{k}_h=-\boldsymbol{\kappa}$.

The electron and hole populations use the same moving grid, cell capacity $M$, and particle weight $w_p$. Let $\mathrm{occ}_a^e$ and $\mathrm{occ}_a^h$ denote the numbers of electrons and holes in cell $\mathcal C_a$. Their occupation fractions and the corresponding occupied-valence count are
\begin{equation}
   F_a=\frac{\mathrm{occ}_a^e}{M},
   \qquad
   G_a=\frac{\mathrm{occ}_a^h}{M},
   \qquad
   \mathrm{occ}_a^v=M-\mathrm{occ}_a^h.
   \label{eq:bipolar-cell-occupations}
\end{equation}
The quantity $\mathrm{occ}_a^v$ represents the occupied valence states in the same discrete normalization; no separate ensemble of valence electrons is stored. With the fixed particle weight defined in Eq.~\eqref{eq:nemc-occupation}, $n_e=w_pN_e$ and $n_h=w_pN_h$, where $N_e$ and $N_h$ are the current numbers of simulated electrons and holes.

During the drift substep, the stored hole coordinate undergoes the same field-driven translation as the electron coordinate and the moving grid,
\begin{equation}
   \boldsymbol{\kappa}_j^\star
   =
   \boldsymbol{\kappa}_j^r-\frac{e}{\hbar}\mathbf{E}\Delta t.
   \label{eq:bipolar-drift}
\end{equation}
The electron and hole cell counts are therefore unchanged during drift. In the physical hole coordinate $\mathbf{k}_h=-\boldsymbol{\kappa}$, Eq.~\eqref{eq:bipolar-drift} gives $\mathbf{k}_h^\star=\mathbf{k}_h^r+(e/\hbar)\mathbf{E}\Delta t$, which is the drift of a positively charged carrier.

During the collision substep, the grid is held fixed.
Intraband transitions of electrons are treated as
described in Section~\ref{sec:nemc}.
A hole transition $\boldsymbol{\kappa}\to
\boldsymbol{\kappa}'$ occurs when a valence electron
scatters from $\boldsymbol{\kappa}'$ into the vacant
state at $\boldsymbol{\kappa}$, leaving a vacancy
at $\boldsymbol{\kappa}'$.
Therefore, the hole transition uses the reversed
valence-electron kernel
$S_{--}(\boldsymbol{\kappa}',\boldsymbol{\kappa})$.
For the Dirac bands and electron--phonon kernels
considered here, the corresponding rates and angular
distributions have the same form as for conduction
electrons when expressed in terms of the positive
energy $\varepsilon=\hbar v_F|\boldsymbol{\kappa}|$.
The acceptance rule in
Eq.~\eqref{eq:nemc-intraband-acceptance} is applied
separately to the electron and hole occupations,
including the same-cell correction when the initial
and final states belong to the same cell.

With the definitions in Eq.~\eqref{eq:FG-definitions},
the generation contribution contains the occupied-valence
factor $1-G$ and the available-conduction factor $1-F$,
while the recombination contribution contains the electron
occupation $F$ and the hole occupation $G$.
In the Monte Carlo construction, the source occupation
factors enter through source sampling, whereas the
final-state factors enter through Pauli acceptance tests.
For a phonon mode $\nu\in\{O,K\}$, the bare interband
rates, before applying these occupation factors, are
\begin{equation}
   \Gamma_\nu^{\mathrm{gen}}(\varepsilon_v)
   =
   \gamma_\nu n_\nu
   [\hbar\omega_\nu-\varepsilon_v]_+,
   \qquad
   \Gamma_\nu^{\mathrm{rec}}(\varepsilon_e)
   =
   \gamma_\nu(n_\nu+1)
   [\hbar\omega_\nu-\varepsilon_e]_+,
   \label{eq:interband-bare-rates}
\end{equation}
where $\varepsilon_v=\hbar v_F|\boldsymbol{\kappa}|$,
$\varepsilon_e=\hbar v_F|\mathbf{k}|$, and
$[x]_+=\max(x,0)$. Generation corresponds to phonon
absorption and recombination to phonon emission.
The prefactors $\gamma_\nu$ and the differential
transition kernels are given in Appendix~\ref{app:phonons}.

For the optical and $K$ modes, the angular probability densities used to sample the interband target state are
\begin{equation}
   p_O(\phi)=\frac{1}{2\pi},
   \qquad
   p_K(\phi)=\frac{1-\cos\phi}{2\pi},
   \qquad
   0\leq\phi<2\pi .
   \label{eq:interband-angular-laws}
\end{equation}
Here $\phi$ is the angle between the initial and final electron-state wave vectors; for a valence-band state, the relevant wave vector is $\boldsymbol{\kappa}$.

Since valence electrons are not explicitly simulated,
generation proposals are constructed from the occupied
valence states in each cell. The cell-averaged bare
generation rates are approximated using a tensor-product
Gauss--Legendre quadrature of order $q_{\mathrm{gen}}$
in each wave-vector direction, giving
$N_q=q_{\mathrm{gen}}^2$ quadrature points
$\boldsymbol{\kappa}_{a\ell}$ with positive normalized
weights $\beta_\ell$, satisfying
$\sum_{\ell=1}^{N_q}\beta_\ell=1$
(see Appendix~\ref{app:nemc}).
For phonon mode $\nu$, the cell-averaged rate is
approximated by
\begin{equation}
   \overline{\Gamma}_{\nu a}^{\mathrm{gen}}
   =
   \sum_{\ell=1}^{N_q}
   \beta_\ell
   \Gamma_\nu^{\mathrm{gen}}
   \!\left(\hbar v_F|\boldsymbol{\kappa}_{a\ell}|\right)
   \simeq
   \frac{1}{|\mathcal C_a|}
   \int_{\mathcal C_a}
   \Gamma_\nu^{\mathrm{gen}}
   \!\left(\hbar v_F|\boldsymbol{\kappa}|\right)
   \,\mathrm{d}\boldsymbol{\kappa}.
   \label{eq:generation-cell-rate}
\end{equation}
The cell-averaged bare generation rates are recomputed
after each drift substep and remain fixed during the
collision substep. However, the occupation-dependent
proposal rates are updated after each accepted
generation or recombination event.
The generation proposal rates are
\begin{equation}
   \Lambda_{\nu a}^{\mathrm{gen}}
   =
   \mathrm{occ}_a^v\,
   \overline{\Gamma}_{\nu a}^{\mathrm{gen}},
   \qquad
   \Lambda^{\mathrm{gen}}
   =
   \sum_{a,\nu}\Lambda_{\nu a}^{\mathrm{gen}} .
   \label{eq:generation-proposal-rate}
\end{equation}
Conditional on a generation event, the source cell, phonon mode, and quadrature point are sampled with probability
\begin{equation}
   \mathbb P(a,\nu,\ell\mid\mathrm{gen})
   =
   \frac{
   \mathrm{occ}_a^v\,\beta_\ell
   \Gamma_\nu^{\mathrm{gen}}(\varepsilon_{a\ell})
   }{
   \Lambda^{\mathrm{gen}}
   },
   \qquad
   \varepsilon_{a\ell}
   =
   \hbar v_F|\boldsymbol{\kappa}_{a\ell}|.
   \label{eq:generation-sampling-law}
\end{equation}
For the selected source cell $a$, phonon mode $\nu$, and quadrature point $\ell$, we take $\boldsymbol{\kappa}_v=\boldsymbol{\kappa}_{a\ell}$ and $\varepsilon_v=\varepsilon_{a\ell}$. After sampling $\phi$ from $p_\nu$, the proposed conduction-electron wave vector is
\begin{equation}
   \mathbf{k}_e^\ast
   =
   \frac{\hbar\omega_\nu-\varepsilon_v}{\hbar v_F}
   \,\mathbf{u}(\theta_v+\phi),
   \qquad
   \mathbf{u}(\theta)=(\cos\theta,\sin\theta),
   \label{eq:generation-target}
\end{equation}
where $\theta_v$ is the polar angle of $\boldsymbol{\kappa}_v$. For $\boldsymbol{\kappa}_v=\mathbf{0}$, a fixed reference direction is used. If $\mathbf{k}_e^\ast$ lies in conduction cell $\mathcal C_b$, the event is accepted with probability
\begin{equation}
   P_{\mathrm{gen}}
   =
   1-F_b
   =
   1-\frac{\mathrm{occ}_b^e}{M}.
   \label{eq:generation-acceptance}
\end{equation}
An accepted event creates an electron at $\mathbf{k}_e^\ast$ and a hole represented at $\boldsymbol{\kappa}_v$, with the simultaneous update
\begin{equation}
   \mathrm{occ}_b^e\leftarrow \mathrm{occ}_b^e+1,
   \qquad
   \mathrm{occ}_a^h\leftarrow \mathrm{occ}_a^h+1.
   \label{eq:generation-update}
\end{equation}
Recombination proposals are generated from the electron ensemble. For an electron $i$ with energy $\varepsilon_i=\hbar v_F|\mathbf{k}_i|$, define
\begin{equation}
   \Gamma_i^{\mathrm{rec}}
   =
   \sum_\nu \Gamma_\nu^{\mathrm{rec}}(\varepsilon_i),
   \qquad
   \Lambda^{\mathrm{rec}}
   =
   \sum_{i=1}^{N_e}\Gamma_i^{\mathrm{rec}}.
   \label{eq:recombination-proposal-rate}
\end{equation}
For a recombination proposal, the electron $i$ and phonon mode $\nu$ are selected with probability $\mathbb P(i,\nu\mid\mathrm{rec})=\Gamma_\nu^{\mathrm{rec}}(\varepsilon_i)/\Lambda^{\mathrm{rec}}$. The electron occupation factor $F$ is included through the sampling of an existing electron.

For the selected electron $i$ and phonon mode $\nu$, after sampling $\phi$ from $p_\nu$, the target valence-state wave vector is
\begin{equation}
   \boldsymbol{\kappa}^\ast
   =
   \frac{\hbar\omega_\nu-\varepsilon_i}{\hbar v_F}
   \,\mathbf{u}(\theta_i+\phi),
   \label{eq:recombination-target}
\end{equation}
where $\theta_i$ is the polar angle of $\mathbf{k}_i$. If $\boldsymbol{\kappa}^\ast$ lies in cell $\mathcal C_b$, the recombination event is accepted with probability
\begin{equation}
   P_{\mathrm{rec}}
   =
   G_b
   =
   \frac{\mathrm{occ}_b^h}{M}.
   \label{eq:recombination-acceptance}
\end{equation}
On acceptance, one hole in $\mathcal C_b$ is selected uniformly and removed together with electron $i$. If the electron belongs to cell $\mathcal C_a$, the simultaneous count update is
\begin{equation}
   \mathrm{occ}_a^e\leftarrow \mathrm{occ}_a^e-1,
   \qquad
   \mathrm{occ}_b^h\leftarrow \mathrm{occ}_b^h-1.
   \label{eq:recombination-update}
\end{equation}
The continuous target $\boldsymbol{\kappa}^\ast$ satisfies the interband scattering kinematics exactly. On acceptance, one of the stored holes in the target cell is selected uniformly and removed; its wave vector $\boldsymbol{\kappa}_j$ generally differs from $\boldsymbol{\kappa}^\ast$. For square cells of side $\Delta k$,
\begin{equation}
   \left|
   \hbar v_F|\boldsymbol{\kappa}_j|
   -
   \hbar v_F|\boldsymbol{\kappa}^\ast|
   \right|
   \leq
   \sqrt{2}\,\hbar v_F\Delta k ,
   \label{eq:recombination-energy-error-bound}
\end{equation}
since both wave vectors belong to the same cell. The eventwise energy discrepancy is therefore $O(\Delta k)$ and vanishes under grid refinement. The associated finite-cell effect is assessed through the grid-refinement results in Section~\ref{sec:results}.

During the collision substep, intraband scattering, generation, and recombination are treated as a single continuous-time event process. Let $\Gamma_{\mathrm{intra}}(\varepsilon)$ denote the total bare intraband electron--phonon rate, and let $\varepsilon_j^h=\hbar v_F|\boldsymbol{\kappa}_j|$. The corresponding ensemble rates are
\begin{equation}
   \Lambda_{\mathrm{intra}}^e
   =
   \sum_{i=1}^{N_e}\Gamma_{\mathrm{intra}}(\varepsilon_i),
   \qquad
   \Lambda_{\mathrm{intra}}^h
   =
   \sum_{j=1}^{N_h}\Gamma_{\mathrm{intra}}(\varepsilon_j^h),
   \label{eq:intraband-ensemble-rates}
\end{equation}
and the total proposal rate is
\begin{equation}
   \Lambda
   =
   \Lambda_{\mathrm{intra}}^e
   +
   \Lambda_{\mathrm{intra}}^h
   +
   \Lambda^{\mathrm{gen}}
   +
   \Lambda^{\mathrm{rec}}.
   \label{eq:bipolar-total-proposal-rate}
\end{equation}
The waiting time to the next proposal is sampled as
\begin{equation}
   \tau
   =
   -\frac{\ln\eta}{\Lambda},
   \qquad
   \eta\sim\mathcal U(0,1).
   \label{eq:bipolar-event-time}
\end{equation}
If $\tau$ lies within the remaining collision interval, the collision clock is advanced by $\tau$ and the event type is selected with probability proportional to its contribution to $\Lambda$. The corresponding source, phonon mode, and target state are then sampled according to the rules described above.

In the NEMC implementations of Refs.~\cite{Romano2015,Coco2021,ZalinyanNastasi2026}, collision times were generated using the standard self-scattering procedure commonly employed in ensemble Monte Carlo methods~\cite{JacoboniReggiani1983}. In the present collision substep, the total proposal rate $\Lambda$ is evaluated for the current state and remains constant until the next proposal. The waiting time can therefore be sampled directly from Eq.~\eqref{eq:bipolar-event-time}, without introducing an additional self-scattering rate. A proposal rejected by the occupation or domain test leaves the state unchanged, while the sampled waiting time is still consumed.

If the sampled waiting time exceeds the remaining duration of the collision substep, no further proposal is generated in that substep. A proposed target outside the computational domain is rejected. The dependence on the finite wave-vector domain is examined together with the other numerical discretization parameters.

\begin{figure}[pos=htbp]
\centering
\includegraphics[width=0.75\linewidth]{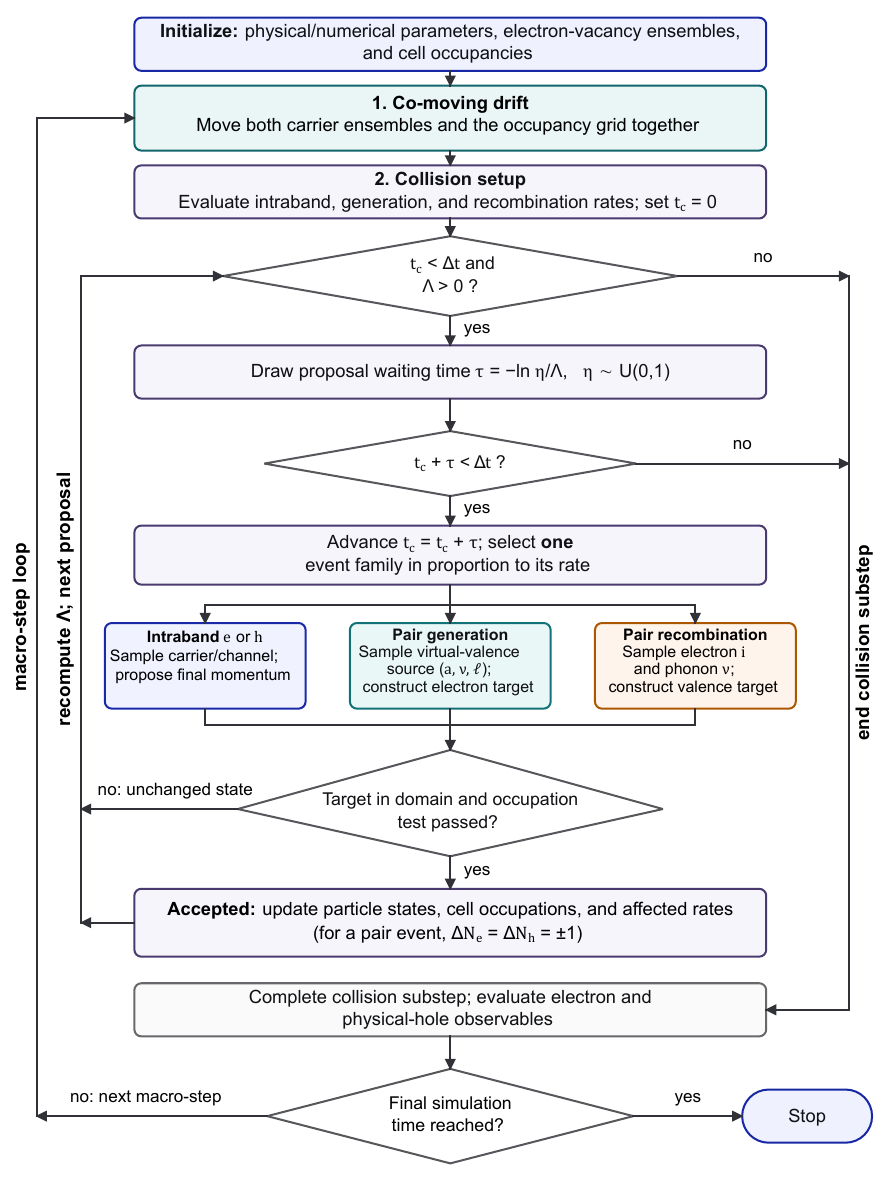}
\caption{Flowchart of the Pauli-consistent bipolar NEMC method.}
\label{fig:nemc-flowchart}
\end{figure}

After each accepted event, the particle states, cell
occupations, and affected proposal rates are updated
before the next waiting time is sampled. Newly generated
particles enter the subsequent rates immediately, while
removed particles no longer contribute.
A change in $\mathrm{occ}_a^h$ also changes the generation
proposal rate of cell $\mathcal C_a$ through
Eq.~\eqref{eq:generation-proposal-rate}.
The complete procedure is illustrated in
Fig.~\ref{fig:nemc-flowchart}.

\section{Numerical results}\label{sec:results}

The numerical results are obtained for spatially homogeneous transport in suspended
monolayer graphene at a lattice temperature of $T=300$ K, starting from the
Fermi--Dirac distributions in Eq.~\eqref{eq:initial-fermi-dirac}. We consider
initial Fermi energies $\varepsilon_F=0$, $0.05$, $0.10$, and $0.20$ eV and
constant electric fields $\mathbf{E}=(E_x,0)$, with
$E_x=0.5$, $1$, $2$, $\ldots$, $10$ kV/cm. Owing to electron--hole
symmetry, only nonnegative values of $\varepsilon_F$ are examined. The main
NEMC and DG simulations extend to $400$ ps, with field-dependent observables
averaged over the interval $300$--$400$ ps.
Throughout the numerical results, $J=J_x$ denotes the
longitudinal component of the total current density.

The NEMC results are compared with an independent deterministic
discontinuous Galerkin (DG) solution of the same bipolar kinetic
model, following Refs.~\cite{MajoranaNastasiRomano2019,NastasiRomano2026}.
NEMC uses continuous particle coordinates and a moving square
occupation grid, whereas DG employs a piecewise-constant
representation on a fixed energy--angle grid. The main numerical
settings are summarized in Table~\ref{tab:main-setup}, and supporting
details of the DG implementation are given in Appendix~\ref{app:dg}.
The target total ensemble size in NEMC is
$N_{\rm tar}=2\times10^6$ particles, with the
initial electron and hole populations determined
by the Fermi energy.
Here $N_{\mathrm{tar}}\simeq N_e^0+N_h^0$,
where $N_e^0$ and $N_h^0$ are the initial electron
and hole particle numbers. At charge neutrality,
$N_e^0\simeq N_h^0\simeq N_p=10^6$.
The initialization procedure for arbitrary
$\varepsilon_F$ is described in
Appendix~\ref{app:nemc}.
All NEMC--DG
runtime comparisons were performed on identical CPU hardware
using the same computational resources per case.

\begin{table}[!th]
\centering
\caption{Main numerical settings for the NEMC--DG comparison.}
\label{tab:main-setup}
\footnotesize
\renewcommand{\arraystretch}{1.12}
\setlength{\tabcolsep}{4pt}
\begin{tabularx}{\textwidth}{@{}>{\raggedright\arraybackslash}p{0.16\textwidth}
                                >{\raggedright\arraybackslash}p{0.44\textwidth}
                                >{\raggedright\arraybackslash}X@{}}
\toprule
Quantity & NEMC & DG \\
\midrule
Phase space
& Continuous particles; moving square occupation grid
& Fixed energy--angle grid \\

Resolution
& $N_k=800$
& $N_E\times N_\theta=500\times72$ \\

Domain
& Square $\mathbf{k}$-domain; $\varepsilon_{\max}=2$ eV
& Radial disk; $\varepsilon_{\max}=2$ eV \\

Particle ensemble
& $N_{\mathrm{tar}}=2\times10^6$; nominal $N_p=10^6$
& -- \\

Cell capacity $M$
& $7239$ at $\varepsilon_F=0$; case-dependent
& -- \\

Time integration
& Pauli-consistent Lagrangian drift--collision update
& Conservative SSP-RK3 \\

Time stepping
& Macro step $\Delta t=2.5$ fs
& Adaptive, CFL-controlled; $\Delta t_{\max}=2.5$ fs \\

Limiter / CFL
& --
& Minmod; $C_{\mathrm{CFL}}=0.35$ \\

NEMC quadrature
& $q_{\mathrm{init}}=4$, $q_{\mathrm{gen}}=2$
& -- \\
\bottomrule
\end{tabularx}
\end{table}

\begin{figure}[pos=htbp]
\centering
\includegraphics[width=0.75\linewidth]{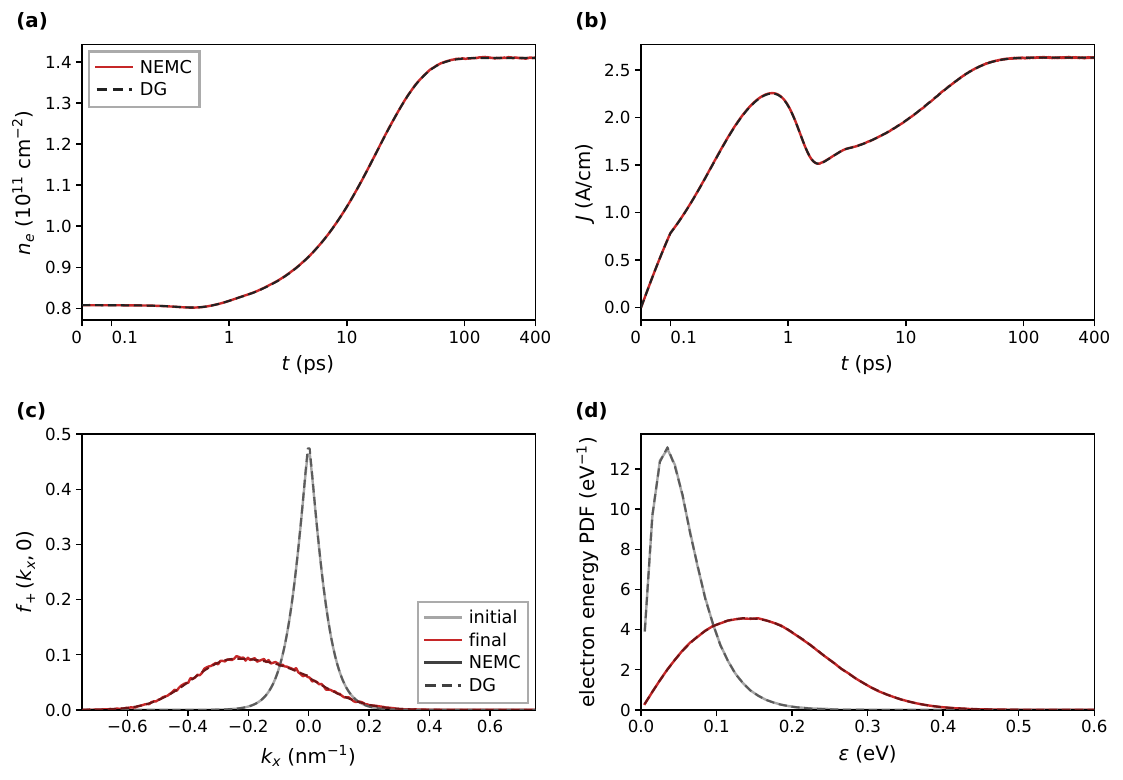}
\caption{NEMC--DG comparison at $\varepsilon_F=0$
and $E_x=2$ kV/cm.
(a) Electron density;
(b) total current density;
(c) electron occupation along $k_y=0$;
(d) electron energy distribution.
Panels (c) and (d) show the initial and final
($t=400$ ps) distributions.}
\label{fig:neutral-case}
\end{figure}

As a representative case, Fig.~\ref{fig:neutral-case}
presents the NEMC and DG results at charge neutrality
($\varepsilon_F=0$) for $E_x=2$ kV/cm. The comparison
includes the time evolution of the transport observables and the
initial and final electron occupation and energy
distributions.

\begin{figure}[pos=htbp]
\centering
\includegraphics[width=0.75\linewidth]{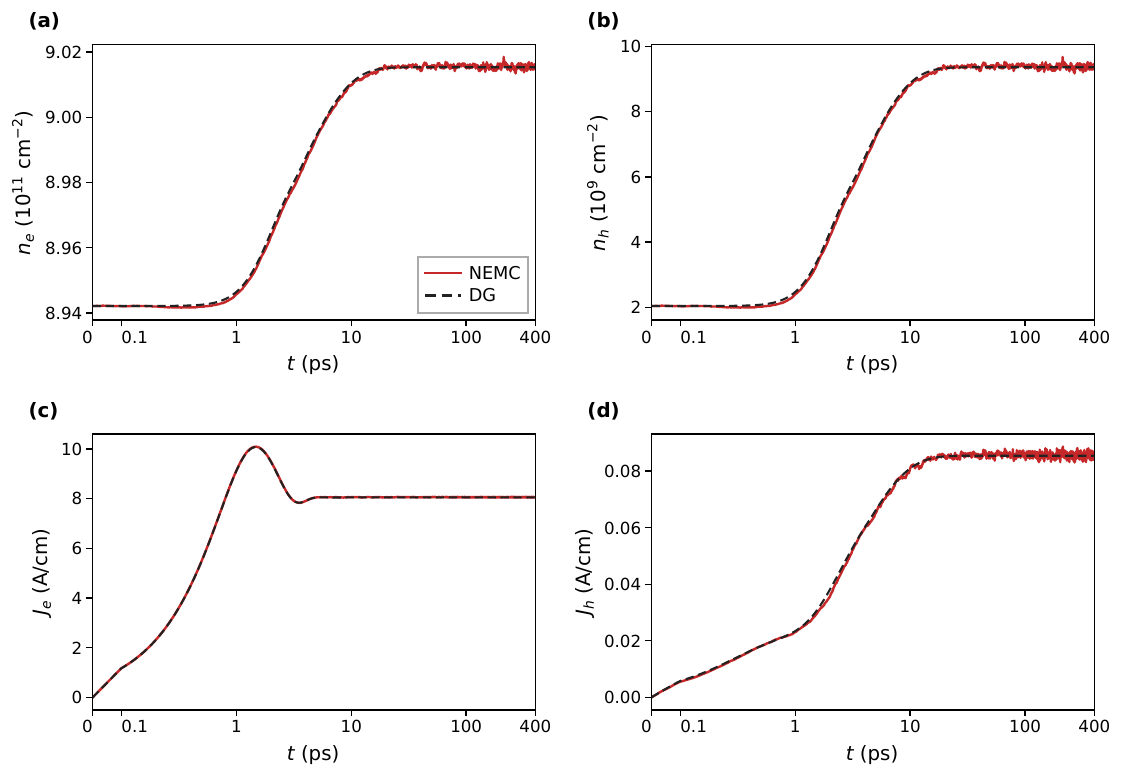}
\caption{NEMC--DG comparison at
$\varepsilon_F=0.10$ eV and $E_x=1$ kV/cm.
(a) Electron density;
(b) hole density;
(c) electron current density;
(d) hole current density.}
\label{fig:bipolar-case}
\end{figure}

As a representative finite-doping case,
Fig.~\ref{fig:bipolar-case} compares the NEMC and DG
results for $\varepsilon_F=0.10$ eV and $E_x=1$ kV/cm,
showing the time evolution of the electron and hole densities
and their separate current contributions.
Additional time-dependent comparisons at
$\varepsilon_F=0.05$ and $0.20$ eV are presented
in Fig.~\ref{fig:additional-transients}.

\begin{figure}[pos=htbp]
\centering
\includegraphics[width=0.75\linewidth]{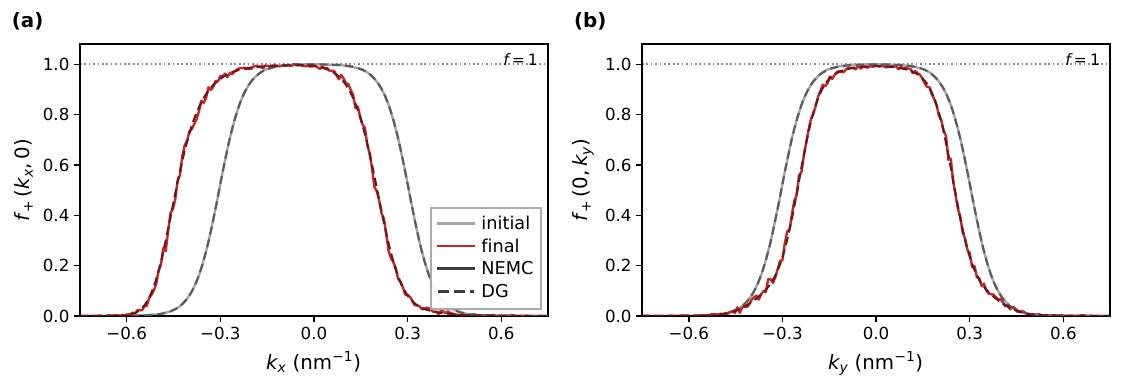}
\caption{NEMC--DG comparison of electron occupation
profiles at $\varepsilon_F=0.20$ eV and
$E_x=0.5$ kV/cm.
(a) Longitudinal profile ($k_y=0$);
(b) transverse profile ($k_x=0$).
Initial and final ($t=400$ ps) distributions are shown.}
\label{fig:occupation-profiles}
\end{figure}

To examine the electron occupation at higher doping,
Fig.~\ref{fig:occupation-profiles} compares the initial
and final occupation profiles along the longitudinal
($k_y=0$) and transverse ($k_x=0$) directions for
$\varepsilon_F=0.20$ eV and $E_x=0.5$ kV/cm.
The upper occupation bound $f=1$, imposed by the
Pauli exclusion principle, is indicated for reference.

\begin{figure}[pos=htbp]
\centering
\includegraphics[width=0.75\linewidth]{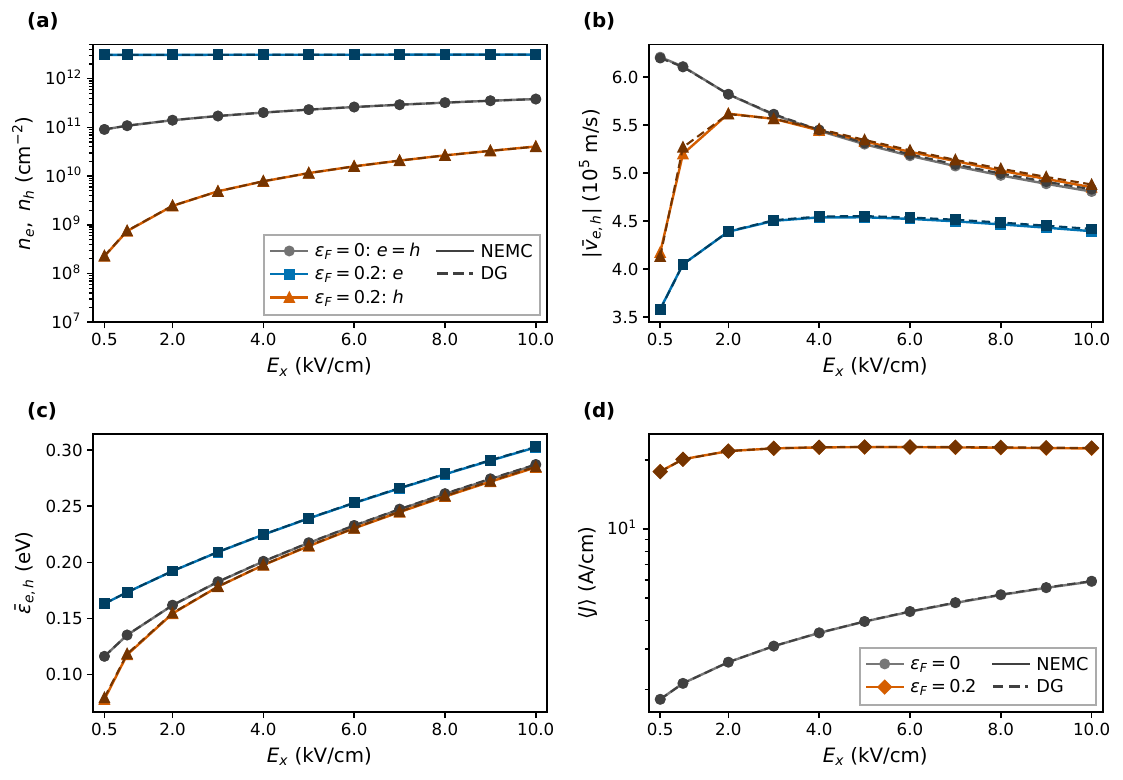}
\caption{Field-dependent NEMC--DG comparison at
$\varepsilon_F=0$ and $0.20$ eV, averaged over
300--400 ps.
(a) Electron and hole densities;
(b) drift-velocity magnitudes;
(c) mean carrier energies;
(d) total current density.}
\label{fig:field-dependent}
\end{figure}

Figure~\ref{fig:field-dependent} compares the
field-dependent NEMC and DG results at
$\varepsilon_F=0$ and $0.20$ eV, averaged over
300--400 ps. At charge neutrality, the electron
and hole observables coincide by symmetry.
The corresponding results for $\varepsilon_F=0.05$
and $0.10$ eV are provided in
Fig.~\ref{fig:intermediate-field}.

\begin{figure}[pos=htbp]
\centering
\includegraphics[width=0.75\linewidth]{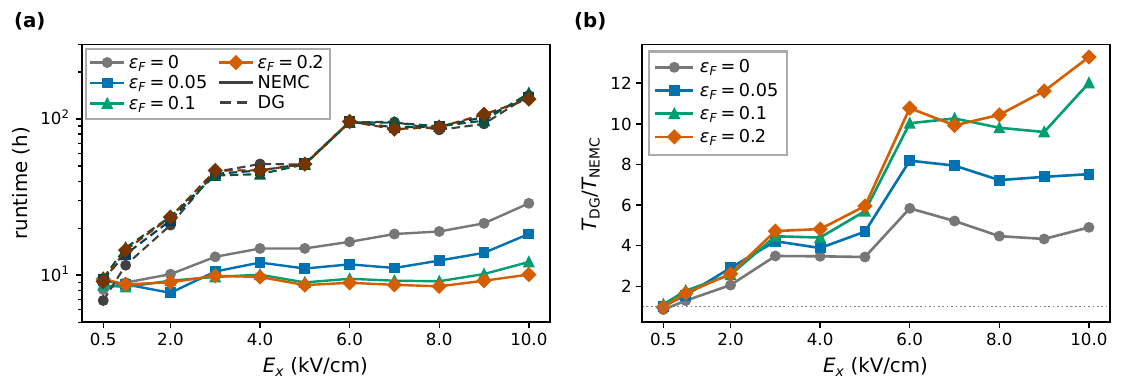}
\caption{Computational cost of the NEMC and DG
simulations for $\varepsilon_F=0$, $0.05$, $0.10$,
and $0.20$ eV.
(a) Wall-clock runtimes;
(b) runtime ratio $T_{\mathrm{DG}}/T_{\mathrm{NEMC}}$
as a function of $E_x$.}
\label{fig:runtime}
\end{figure}

Figure~\ref{fig:runtime} presents the measured runtimes
of the 400 ps NEMC and DG simulations and the
field-dependent DG-to-NEMC runtime ratios for all four
initial Fermi energies.

\begin{figure}[pos=htbp]
\centering
\includegraphics[width=0.75\linewidth]{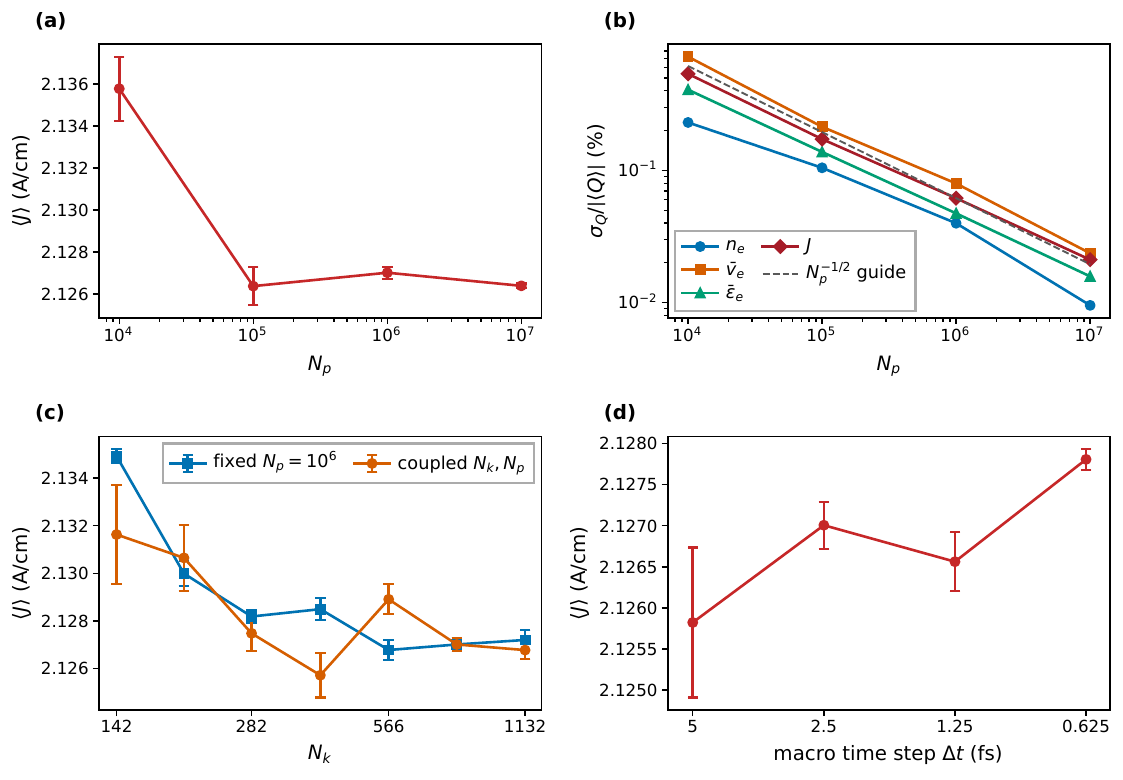}
\caption{NEMC numerical sensitivity at
$\varepsilon_F=0$ and $E_x=1$ kV/cm.
(a) Late-time mean current versus ensemble size at $N_k=800$;
(b) relative temporal standard deviations versus ensemble size;
(c) late-time mean current versus occupation-grid resolution
for fixed $N_p=10^6$ and coupled $(N_k,N_p)$ refinement,
with $N_p$ increasing from $31250$ at $N_k=142$
to $2\times10^6$ at $N_k=1132$;
(d) late-time mean current versus macro timestep at
$N_k=800$ and $N_p=10^6$.
All simulations extend to 200 ps, with late-time statistics
evaluated over 160--200 ps.}
\label{fig:nemc-sensitivity}
\end{figure}

The effect of the ensemble size was examined at fixed
$N_k=800$ for $\varepsilon_F=0$ and $E_x=1$ kV/cm,
with $N_p$ ranging from $10^4$ to $10^7$.
Figure~\ref{fig:nemc-sensitivity}(a) compares the
late-time mean current densities, while panel (b)
shows the relative temporal standard deviations
of the transport observables. The $N_p^{-1/2}$
dependence is included as a reference scaling for
the statistical fluctuations. Quantitative results are reported in
Table~\ref{tab:np-sensitivity}.
Here and in the following numerical-sensitivity tests,
unless stated otherwise, simulations are run for 200 ps,
with late-time statistics evaluated over 160--200 ps.

The sensitivity to the occupation-grid resolution was
examined by increasing $N_k$ along two refinement paths:
at fixed $N_p=10^6$ and with $N_p$ increased proportionally
to $N_k^2$ to maintain approximately constant particle
statistics per cell. The coupled refinement uses
$N_k=142$, $200$, $282$, $400$, $566$, $800$, and $1132$,
with corresponding
$N_p=31250$, $62500$, $125000$, $250000$,
$500000$, $10^6$, and $2\times10^6$, respectively.
Figure~\ref{fig:nemc-sensitivity}(c) compares the
late-time current for these refinements.
Additional NEMC refinement data are given in
Table~\ref{tab:d-nemc}.
Two-dimensional NEMC--DG occupation comparisons
are presented in Fig.~\ref{fig:occupation-maps}.
The DG mesh-refinement results are given in
Fig.~\ref{fig:dg-refinement} and
Table~\ref{tab:d-dg}.

The sensitivity to the NEMC macro timestep was examined
at fixed $N_k=800$ and $N_p=10^6$ for
$\varepsilon_F=0$ and $E_x=1$ kV/cm
(Fig.~\ref{fig:nemc-sensitivity}(d)).
Reducing $\Delta t$ from $1.25$ to $0.625$ fs changes
the late-time mean current by less than $0.1\%$,
while approximately doubling the runtime.
Results for the additional test cases are reported
in Table~\ref{tab:d-nemc}.

The sensitivity to the generation-quadrature order
$q_{\mathrm{gen}}$ was examined for three representative
cases. The corresponding results are reported in
Table~\ref{tab:d-gen-quadrature}.

\section{Discussion}
\label{sec:discussion}

The NEMC and DG results show very good agreement across the
investigated electric fields and initial Fermi energies, both
for the time evolution and for the steady-state transport
observables. The agreement extends to the electron and hole
populations, drift velocities, mean energies, currents, and
carrier distributions. While the figures illustrate
representative transport conditions, the quantitative
comparison covers all $44$ nonzero-field cases.
The maximum absolute relative differences between the
$300$--$400$ ps NEMC and DG means, normalized by the
corresponding DG values, are $0.13\%$ and $0.97\%$ for
electron and hole densities, $0.53\%$ and $1.30\%$ for
their drift velocities, and $0.17\%$ and $0.55\%$ for
their mean energies, respectively. The maximum difference
in the total current is $0.54\%$, at
$\varepsilon_F=0.20$ eV and $E_x=10$ kV/cm, while the
largest difference among all these observables occurs
for the minority-hole drift velocity at
$\varepsilon_F=0.20$ eV and $E_x=0.5$ kV/cm.
These differences quantify the agreement between two
numerical approximations rather than the absolute
error of NEMC, since the DG solution itself depends
on numerical discretization, including phase-space
grid resolution and time integration, as examined
in the mesh-refinement results of
Fig.~\ref{fig:dg-refinement} and
Table~\ref{tab:d-dg}.
As the two methods employ independent numerical
implementations with different phase-space representations
and time-integration schemes, their agreement provides
mutual numerical cross-validation of the bipolar transport
formulation within the considered kinetic model.

The proposed bipolar NEMC formulation provides a
computationally practical particle-based description
of coupled electron and hole transport.
Unlike DG, which evolves discretized occupation
distributions, NEMC explicitly represents individual
carrier states and scattering events. This provides
direct access to microscopic carrier dynamics and
enables event-resolved investigations that are less
directly accessible with distribution-based
deterministic methods.

The numerical-sensitivity studies provide additional
support for the reliability of the NEMC results.
At $\varepsilon_F=0$ and $E_x=1$ kV/cm, increasing
$N_p$ from $10^4$ to $10^7$ reduces the relative
block SEM of the steady-state mean current from
$0.0714\%$ to $0.00534\%$. Across the six refinement
cases, the final coupled NEMC grid--particle refinement
changes the current by at most $0.11\%$.
Reducing the NEMC macro timestep from $1.25$
to $0.625$ fs changes the current by at most $0.0854\%$.

The generation-quadrature sensitivity is more
pronounced for the minority-carrier dynamics at high
doping. At $\varepsilon_F=0.20$ eV and $E_x=2$ kV/cm,
increasing $q_{\rm gen}$ from $2$ to $4$ changes the
accepted generation rate by $1.23\%$ and the
steady-state hole density by $0.85\%$, whereas the
total current changes by only $0.0027\%$.
In this strongly electron-dominated regime, the
initial ensemble contains $1{,}999{,}145$ electrons
and only $28$ holes. Although generation increases
the hole population during the simulation, the
steady-state hole density remains below $0.08\%$
of the electron density. Consequently, the
minority-carrier contribution to the total current
is small, despite the greater numerical sensitivity
of the hole population and generation rate.
This behavior is consistent with the earlier DG
study of Ref.~\cite{MajoranaNastasiRomano2019},
which showed that, at $\varepsilon_F=0.20$ eV,
neglecting interband scattering has little effect
on the overall transport, although it substantially
affects the hole density. These results also show
that the required numerical resolution depends on
the observable of interest: accurate total current
alone does not necessarily imply comparable accuracy
for minority-carrier populations and interband
transition rates.

The computational-cost comparison in
Fig.~\ref{fig:runtime} demonstrates favorable
computational performance of NEMC in the investigated
transport regimes. Of the $44$ nonzero-field cases,
NEMC requires less computation time in $42$.
At charge neutrality and $E_x=10$ kV/cm, DG
requires approximately $4.9$ times the NEMC runtime.
This ratio reaches $13.3$ at
$\varepsilon_F=0.20$ eV and the same field,
where the recorded runtimes are $134.1$ h for DG
and $10.1$ h for NEMC. An additional computational
advantage arises from the possibility of selecting
the particle ensemble according to the required
accuracy of the observables. In the ensemble-size
study at $\varepsilon_F=0$ and $E_x=1$ kV/cm,
the steady-state mean currents obtained with
$N_p=10^5$ and $10^7$ agree to the reported
precision, while the smaller ensemble requires
approximately $46$ times less computational time.
This demonstrates that the computational cost of
NEMC can be reduced by more than an order of
magnitude for selected mean transport observables,
without requiring the largest particle ensembles.

The two methods have different computational
requirements. NEMC evolves continuous particle
states with a prescribed macro timestep, and
its cost depends on the particle ensemble and
scattering-event statistics. DG is free from
Monte Carlo sampling noise, but its computational
cost depends on the phase-space resolution
and the adaptive timestep restricted by the
transport and collision rates. The measured
runtime advantage of NEMC is therefore specific
to the present implementations and numerical
settings, rather than a general comparison
at identical prescribed accuracy.

In addition to statistical fluctuations, the discrete
occupation-grid representation in NEMC can introduce
weak numerical oscillations in time-dependent observables.
As investigated in Ref.~\cite{ZalinyanNastasi2026},
for electron--phonon-only transport these grid-locked
oscillations become distinguishable in the low-noise
regime reached with particle ensembles exceeding
approximately $10^7$. The present study focuses on
the overall time evolution and steady-state mean
transport observables; the analysis of such
small-amplitude temporal oscillations is therefore
outside its scope.

Although the present numerical investigation considers
electron--phonon scattering in graphene, the underlying
Pauli-consistent bipolar NEMC construction is not
intrinsically restricted to this collision model or
material. Previous work developed the treatment of
electron--electron scattering within NEMC, including
theoretical analysis of interband collision processes,
while the numerical simulations considered unipolar
transport \cite{NastasiRomano2023}. The NEMC treatment
of electron--electron scattering was subsequently
further developed in Ref.~\cite{ZalinyanNastasi2026}.
These developments provide a basis for extending the
present bipolar formulation to include carrier--carrier
interactions. Extensions to other semiconductor systems
would require appropriate modifications of the
dispersion relations, carrier-state representation,
scattering kernels, and interband transition rules.
Further developments may also include nonequilibrium
phonon and thermal effects, as well as spatially
inhomogeneous transport. The latter would require
the consistent treatment of real-space evolution,
boundary conditions, and, where appropriate,
self-consistent electric fields, as considered
in deterministic graphene-device transport in
Ref.~\cite{NastasiRomano2026}. These extensions
remain subjects for future research.

\section{Conclusions}
\label{sec:conclusions}

We have extended the Pauli-consistent ensemble Monte
Carlo method to bipolar charge transport in graphene,
incorporating phonon-assisted electron--hole generation
and recombination. The proposed formulation enables
the simulation of coupled electron and hole populations
while preserving the Pauli occupation bounds and net
charge, and retaining the microscopic particle-based
description of carrier dynamics.

The numerical assessment demonstrates very good
agreement with an independent discontinuous Galerkin
solution for the time evolution, carrier distributions,
and steady-state transport observables. Together with
the numerical-sensitivity and refinement studies,
these results support the reliability of the method
in the investigated regimes. The computational
comparisons also demonstrate its favorable performance,
showing that a microscopic particle-based approach
can provide accurate transport results at competitive
computational cost. Beyond the electron--phonon model
considered here, the bipolar NEMC framework provides
a basis for incorporating additional interactions,
including carrier--carrier scattering, and for further
development toward other semiconductor materials.

\section*{Acknowledgements}

This research was supported by the Higher Education
and Science Committee of MESCS RA (Research Project
No.~25YR-1C015).

The main numerical simulations reported in this work
were carried out on the YSU Supercomputer, which
enabled long-time simulations and parallel calculations.

The author GN acknowledges the support from
GNFM - INdAM and from research fund of the
University of Enna ``Kore'', Italy.

\appendix
\numberwithin{equation}{section}
\counterwithin{table}{section}
\counterwithin{figure}{section}
\renewcommand{\theHequation}{\thesection.\arabic{equation}}
\renewcommand{\theHtable}{\thesection.\arabic{table}}
\renewcommand{\theHfigure}{\thesection.\arabic{figure}}

\section{Electron--phonon scattering model and parameters}
\label{app:phonons}

The electron--phonon transition kernels, scattering rates, angular laws, and material parameters used in the NEMC and DG calculations are summarized here. The effective acoustic parameters are taken from Ref.~\cite{Kaasbjerg2012UnravelingTheAcousticElectron}, while the optical and intervalley parameters are taken from Ref.~\cite{Borysenko2010FirstPrinciplesAnalysisOf}.

In $S_{ss'}(\mathbf{k},\mathbf{k}')$, $(s,\mathbf{k})$ denotes the initial state and $(s',\mathbf{k}')$ the final state. Let $\phi$ be the angle between the initial and final electron-state wave vectors. The total transition kernel is
\begin{equation}
S_{ss'}(\mathbf{k},\mathbf{k}')
=
S_{ss'}^{\mathrm{ac}}(\mathbf{k},\mathbf{k}')
+
\sum_{\nu\in\{O,K\}}
S_{ss'}^\nu(\mathbf{k},\mathbf{k}').
\label{eq:app-total-phonon-kernel}
\end{equation}

The effective acoustic channel is treated in the elastic, equipartition approximation,
\begin{equation}
S_{ss'}^{\mathrm{ac}}(\mathbf{k},\mathbf{k}')
=
\delta_{ss'}
\frac{\pi D_{\mathrm{ac}}^2 k_BT}
     {(2\pi)^2\,2\hbar\rho_m v_p^2}
(1+\cos\phi)\,
\delta\!\left(
\varepsilon_{s'}(\mathbf{k}')
-\varepsilon_s(\mathbf{k})
\right),
\label{eq:app-acoustic-kernel}
\end{equation}
where $D_{\mathrm{ac}}$ is the acoustic deformation potential, $\rho_m$ is the graphene areal mass density, and $v_p$ is the sound velocity. The factor $\delta_{ss'}$ restricts this channel to intraband transitions.

For the dispersionless optical and intervalley modes, writing
$\Delta\varepsilon_{ss'}=
\varepsilon_{s'}(\mathbf{k}')-\varepsilon_s(\mathbf{k})$,
the transition kernels are
\begin{equation}
S_{ss'}^\nu(\mathbf{k},\mathbf{k}')
=
\mathcal K_\nu(\phi)
\left[
(n_\nu+1)
\delta(\Delta\varepsilon_{ss'}+\hbar\omega_\nu)
+
n_\nu
\delta(\Delta\varepsilon_{ss'}-\hbar\omega_\nu)
\right],
\qquad
\nu\in\{O,K\},
\label{eq:app-inelastic-kernel}
\end{equation}
with
\begin{equation}
\mathcal K_O(\phi)
=
\frac{2\pi D_O^2}
     {(2\pi)^2\rho_m\omega_O},
\qquad
\mathcal K_K(\phi)
=
\frac{2\pi D_K^2}
     {(2\pi)^2\rho_m\omega_K}
(1-\cos\phi).
\label{eq:app-inelastic-amplitudes}
\end{equation}
Here $n_\nu$ is the equilibrium Bose occupation defined in
Eq.~\eqref{eq:phonon-occupation}. The terms proportional to
$n_\nu+1$ and $n_\nu$ describe phonon emission and absorption,
respectively. The effective $O$ channel combines the LO and TO
contributions and is isotropic, while the $K$ channel has the
$1-\cos\phi$ angular dependence.

For interband transitions, $\phi$ is defined between the conduction- and valence-electron wave vectors. In the bipolar representation, the valence-band wave vector entering the kernel is $\boldsymbol{\kappa}$.

The bare scattering rates are obtained by integrating the transition kernels over the final states before the occupation factors are applied. For the linear Dirac dispersion, the resulting rate prefactors are
\begin{equation}
\gamma_{\mathrm{ac}}
=
\frac{D_{\mathrm{ac}}^2 k_BT}
     {4\hbar^3\rho_m v_p^2v_F^2},
\qquad
\gamma_\nu
=
\frac{D_\nu^2}
     {\rho_m\omega_\nu\hbar^2v_F^2},
\qquad
\nu\in\{O,K\}.
\label{eq:app-rate-prefactors}
\end{equation}

Table~\ref{tab:app-phonon-rates} summarizes the positive target energy and the corresponding bare rate for each scattering process. For intraband events, $\varepsilon$ denotes the positive electron or hole energy; for generation and recombination it denotes the source valence depth $\varepsilon_v$ and source conduction energy $\varepsilon_e$, respectively.

\begin{table}[pos=htbp]
\centering
\caption{Positive target energies and bare electron--phonon scattering rates. Here $\nu\in\{O,K\}$ and $[x]_+=\max(x,0)$.}
\label{tab:app-phonon-rates}
\begin{tabular}{lll}
\hline
Process & Target energy & Bare rate \\
\hline
Acoustic intraband
& $\varepsilon$
& $\gamma_{\mathrm{ac}}\varepsilon$ \\[2pt]

Intraband emission
& $\varepsilon-\hbar\omega_\nu$
& $\gamma_\nu(n_\nu+1)[\varepsilon-\hbar\omega_\nu]_+$ \\[2pt]

Intraband absorption
& $\varepsilon+\hbar\omega_\nu$
& $\gamma_\nu n_\nu(\varepsilon+\hbar\omega_\nu)$ \\[2pt]

Interband generation
& $\hbar\omega_\nu-\varepsilon_v$
& $\gamma_\nu n_\nu[\hbar\omega_\nu-\varepsilon_v]_+$ \\[2pt]

Interband recombination
& $\hbar\omega_\nu-\varepsilon_e$
& $\gamma_\nu(n_\nu+1)[\hbar\omega_\nu-\varepsilon_e]_+$ \\
\hline
\end{tabular}
\end{table}

Intraband emission is allowed only for
$\varepsilon>\hbar\omega_\nu$, whereas interband generation and
recombination require the corresponding source energy to satisfy
$\varepsilon_v<\hbar\omega_\nu$ and
$\varepsilon_e<\hbar\omega_\nu$, respectively. The total bare
intraband electron--phonon rate is therefore
\begin{equation}
\Gamma_{\mathrm{intra}}(\varepsilon)
=
\gamma_{\mathrm{ac}}\varepsilon
+
\sum_{\nu\in\{O,K\}}
\gamma_\nu
\left[
(n_\nu+1)[\varepsilon-\hbar\omega_\nu]_+
+
n_\nu(\varepsilon+\hbar\omega_\nu)
\right].
\label{eq:app-total-intraband-rate}
\end{equation}

The normalized angular probability densities are
\begin{equation}
p_{\mathrm{ac}}(\phi)
=
\frac{1+\cos\phi}{2\pi},
\qquad
p_O(\phi)
=
\frac{1}{2\pi},
\qquad
p_K(\phi)
=
\frac{1-\cos\phi}{2\pi},
\qquad
0\leq\phi<2\pi.
\label{eq:app-angular-densities}
\end{equation}
These angular laws are used to sample the final-state direction for the corresponding scattering process.

All rates given above are bare transition rates. The corresponding occupation factors enter separately through the collision terms; their implementation in the bipolar NEMC method is described in Section~\ref{sec:bipolar}.

The material parameters used in the calculations are listed in
Table~\ref{tab:app-material-parameters}. The acoustic parameters are
taken from Ref.~\cite{Kaasbjerg2012UnravelingTheAcousticElectron},
while the optical and intervalley parameters are taken from
Ref.~\cite{Borysenko2010FirstPrinciplesAnalysisOf}.

\begin{table}[pos=htbp]
\centering
\caption{Parameters of the effective electron--phonon model.}
\label{tab:app-material-parameters}
\begin{tabular}{llll}
\hline
Quantity & Symbol & Value & Source \\
\hline
Fermi velocity
& $v_F$
& $1.0\times10^6\,\mathrm{m/s}$
& \cite{Kaasbjerg2012UnravelingTheAcousticElectron} \\

Sound velocity
& $v_p$
& $2.0\times10^4\,\mathrm{m/s}$
& \cite{Kaasbjerg2012UnravelingTheAcousticElectron} \\

Areal mass density
& $\rho_m$
& $7.6\times10^{-7}\,\mathrm{kg/m^2}$
& \cite{Kaasbjerg2012UnravelingTheAcousticElectron} \\

Acoustic deformation potential
& $D_{\mathrm{ac}}$
& $6.8\,\mathrm{eV}$
& \cite{Kaasbjerg2012UnravelingTheAcousticElectron} \\

Optical phonon energy
& $\hbar\omega_O$
& $164.6\,\mathrm{meV}$
& \cite{Borysenko2010FirstPrinciplesAnalysisOf} \\

Optical deformation potential
& $D_O$
& $1.0\times10^{11}\,\mathrm{eV/m}$
& \cite{Borysenko2010FirstPrinciplesAnalysisOf} \\

Intervalley phonon energy
& $\hbar\omega_K$
& $124\,\mathrm{meV}$
& \cite{Borysenko2010FirstPrinciplesAnalysisOf} \\

Intervalley deformation potential
& $D_K$
& $3.5\times10^{10}\,\mathrm{eV/m}$
& \cite{Borysenko2010FirstPrinciplesAnalysisOf} \\
\hline
\end{tabular}
\end{table}

The phonon bath is kept at the lattice temperature $T$ throughout the simulations. Spin and valley degeneracies enter the carrier densities and particle weight through $g_s=g_v=2$, as defined in Section~\ref{sec:physical}, but do not introduce an additional multiplicative factor in the single-particle transition rates above.

\section{Supporting numerical details of the NEMC construction}
\label{app:nemc}

This appendix summarizes the numerical details required for the cell quadrature, initialization of the electron and hole populations, and treatment of the finite wave-vector domain.

Consider a square cell $\mathcal C_a$ with centre $\mathbf c_a$ and side $\Delta k$. Let $\{z_r,\omega_r\}_{r=1}^{q}$ be the nodes and weights of the $q$-point Gauss--Legendre rule on $[-1,1]$. The tensor-product nodes and normalized weights are
\begin{equation}
\boldsymbol{\xi}_{a,rs}
=
\mathbf c_a
+
\frac{\Delta k}{2}(z_r,z_s),
\qquad
\beta_{rs}
=
\frac{\omega_r\omega_s}{4},
\qquad
\sum_{r,s=1}^{q}\beta_{rs}=1.
\label{eq:app-cell-quadrature}
\end{equation}
Accordingly, the cell average of a function $\psi$ is approximated by
\begin{equation}
\langle\psi\rangle_{a,q}
=
\sum_{r,s=1}^{q}
\beta_{rs}\psi(\boldsymbol{\xi}_{a,rs})
\simeq
\frac{1}{(\Delta k)^2}
\int_{\mathcal C_a}
\psi(\mathbf k)\,\mathrm d^2\mathbf k.
\label{eq:app-quadrature-average}
\end{equation}
Separate orders $q_{\mathrm{init}}$ and $q_{\mathrm{gen}}$ are used for initialization and generation-rate evaluation, respectively; for the generation quadrature, $N_q=q_{\mathrm{gen}}^2$.

The initial electron and hole cell occupations are obtained from the Fermi--Dirac distributions through
\begin{equation}
\overline F_a^0
=
\langle F(0,\cdot)\rangle_{a,q_{\mathrm{init}}},
\qquad
\overline G_a^0
=
\langle G(0,\cdot)\rangle_{a,q_{\mathrm{init}}}.
\label{eq:app-initial-cell-averages}
\end{equation}
Because the quadrature weights are positive and normalized,
$0\leq \overline F_a^0,\overline G_a^0\leq1$.

For a target initial total number of simulated particles
$N_{\mathrm{tar}}$, the common cell capacity is chosen as
\begin{equation}
M
=
\max\left\{
1,\,
\operatorname{nint}\!\left(
\frac{N_{\mathrm{tar}}}
{\sum_a\left(\overline F_a^0+\overline G_a^0\right)}
\right)
\right\},
\label{eq:app-capacity-choice}
\end{equation}
where $\operatorname{nint}$ denotes rounding to the nearest integer.
Once selected, $M$ and the corresponding particle weight $w_p$
remain fixed throughout the simulation and are the same for the
electron and hole populations.

For each species $\alpha\in\{e,h\}$, with
$\overline u_a^e=\overline F_a^0$ and
$\overline u_a^h=\overline G_a^0$, the initial particle number is
\[
N_\alpha^0
=
\operatorname{nint}\!\left(
M\sum_a\overline u_a^\alpha
\right).
\]
The values $M\overline u_a^\alpha$ are then converted to integer
cell counts by a largest-remainder assignment such that
$\sum_a\mathrm{occ}_a^\alpha(0)=N_\alpha^0$ and
$0\leq\mathrm{occ}_a^\alpha(0)\leq M$. The corresponding cellwise
rounding error obeys
\begin{equation}
\left|
\frac{\mathrm{occ}_a^\alpha(0)}{M}
-\overline u_a^\alpha
\right|
\leq
\frac{1}{M}.
\label{eq:app-cell-rounding-error}
\end{equation}

The corresponding initial densities are
$n_e^0=w_pN_e^0$ and $n_h^0=w_pN_h^0$.
For a given Fermi energy, the electron and hole populations are
determined separately from $\overline F_a^0$ and $\overline G_a^0$,
while the same $M$ and $w_p$ are used for both species.

At charge neutrality, $\varepsilon_F=0$, the initial cell averages satisfy
$\overline F_a^0=\overline G_a^0$. The same integer count assignment is
therefore used for both populations, giving
$\mathrm{occ}_a^e(0)=\mathrm{occ}_a^h(0)$ in every cell and exact initial
charge neutrality.

Once the cell counts are fixed, the particle wave vectors are sampled
uniformly within their assigned cells, consistently with the piecewise
constant cell representation. At charge neutrality, the same within-cell
samples are used for the electron and stored-hole populations.

The same cell quadrature, with order $q_{\mathrm{gen}}$, is used to
evaluate the cell-averaged bare generation rates entering
Section~\ref{sec:bipolar}. After each drift step, the quadrature-point
positions and the corresponding cell-averaged generation rates are
updated for the displaced grid.

The simulations are performed on a finite square wave-vector domain.
The occupation grid undergoes the common Lagrangian translation
described in Section~\ref{sec:nemc}. If needed for numerical storage,
its indexing is shifted by an integer number of cells without changing
the represented physical state or particle numbers and without
interpolation. The computational domain is chosen large enough to
contain the relevant electron and hole distributions and the
valence-state region contributing to interband generation.

\section{Supporting details of the DG method}
\label{app:dg}

For an independent deterministic comparison, we use a discontinuous Galerkin (DG) solver for the spatially homogeneous bipolar Boltzmann equations, following the approach of Ref.~\cite{MajoranaNastasiRomano2019}. The DG calculations employ the same electron--phonon collision model and physical parameters as the NEMC simulations, including intraband scattering and phonon-assisted interband generation and recombination. Only the numerical implementation details relevant to the present comparisons are summarized below.

The DG solver uses a uniform grid in positive energy
$\varepsilon=\hbar v_F|\mathbf{k}|$ and polar angle
$\theta\in[0,2\pi)$, with $N_E$ energy cells and $N_\theta$
angular cells.
The initial DG cell averages are evaluated
using a 16-point Gauss--Legendre quadrature in energy.
The conduction-electron and valence-vacancy
occupations are represented by piecewise-constant ($P^0$)
cell averages, with the appropriate momentum-space integration
weights. Unlike the modified-polar discretization of
Ref.~\cite{MajoranaNastasiRomano2019}, the present implementation
uses a uniform energy partition. The electric-field transport
term is discretized conservatively using upwind numerical
fluxes with a minmod-limited reconstruction.

Time integration is performed using an adaptive third-order
strong-stability-preserving Runge--Kutta (SSP-RK3) scheme.
The trial internal time step is selected according to
\begin{equation}
\Delta t_{\mathrm{DG}} =
\min\left\{
\Delta t_{\max},
\frac{C_{\mathrm{CFL}}}{\Lambda_{\max}},
\Delta t_{\mathrm{out}}
\right\},
\end{equation}
where $\Lambda_{\max}$ is a bound incorporating the
electric-field transport and electron--phonon collision
rates, and $\Delta t_{\mathrm{out}}$ is the remaining time
to the next output. The field-flux reconstruction is
bounded, and the occupation constraints $0\leq F,G\leq 1$
are checked at each Runge--Kutta stage with a roundoff
tolerance. A trial step with a violation beyond this
tolerance is rejected and repeated with a reduced time
step. Roundoff-level excursions are removed by projection
only after an admissible step has been obtained.

The energy domain is truncated at $\varepsilon_{\max}$,
with periodicity in $\theta$, vanishing radial flux at
$\varepsilon=0$, and zero exterior occupation at
$\varepsilon_{\max}$. The sensitivity to the finite energy cutoff
is examined in Appendix~\ref{app:additional}.
Electron and hole densities, mean velocities, mean
energies, and total current are obtained from the
discrete distributions using the definitions in
Section~\ref{sec:physical}.

\section{Additional numerical results}
\label{app:additional}

\begin{figure}[pos=htbp]
\centering
\includegraphics[width=0.75\linewidth]{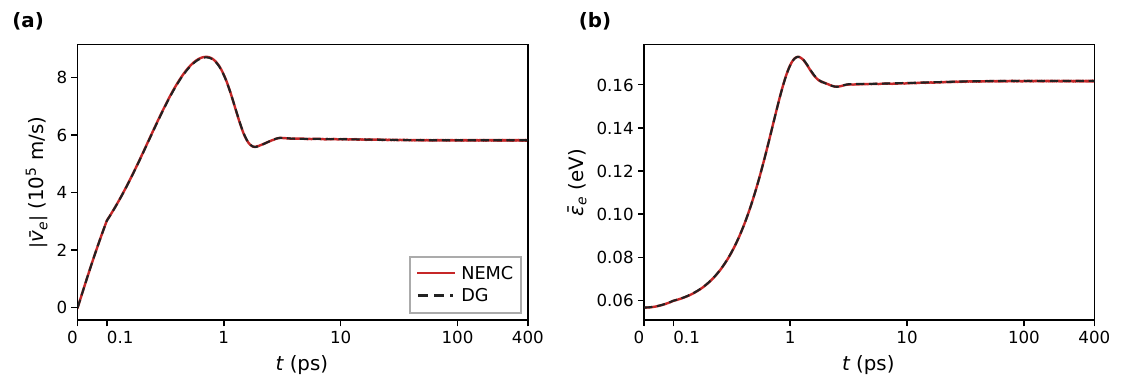}
\caption{NEMC--DG comparison of time-dependent observables at charge neutrality
($\varepsilon_F=0$) and $E_x=2$ kV/cm.
(a) Electron drift-velocity magnitude;
(b) mean electron energy.}
\label{fig:neutral-additional}
\end{figure}

\begin{figure}[pos=htbp]
\centering
\includegraphics[width=0.75\linewidth]{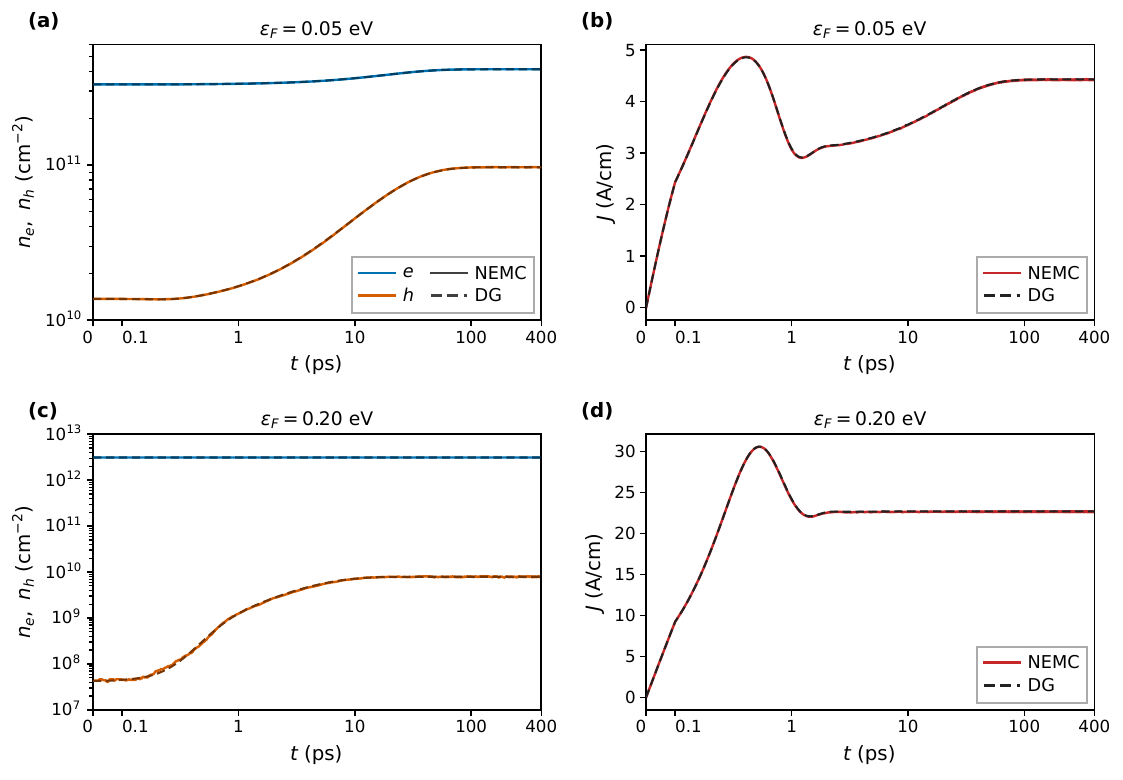}
\caption{NEMC--DG comparison of time-dependent observables at $E_x=4$ kV/cm for
$\varepsilon_F=0.05$ eV (a,b) and $0.20$ eV (c,d).
(a,c) Electron and hole densities;
(b,d) total current density.}
\label{fig:additional-transients}
\end{figure}

\begin{figure}[pos=htbp]
\centering
\includegraphics[width=0.75\linewidth]{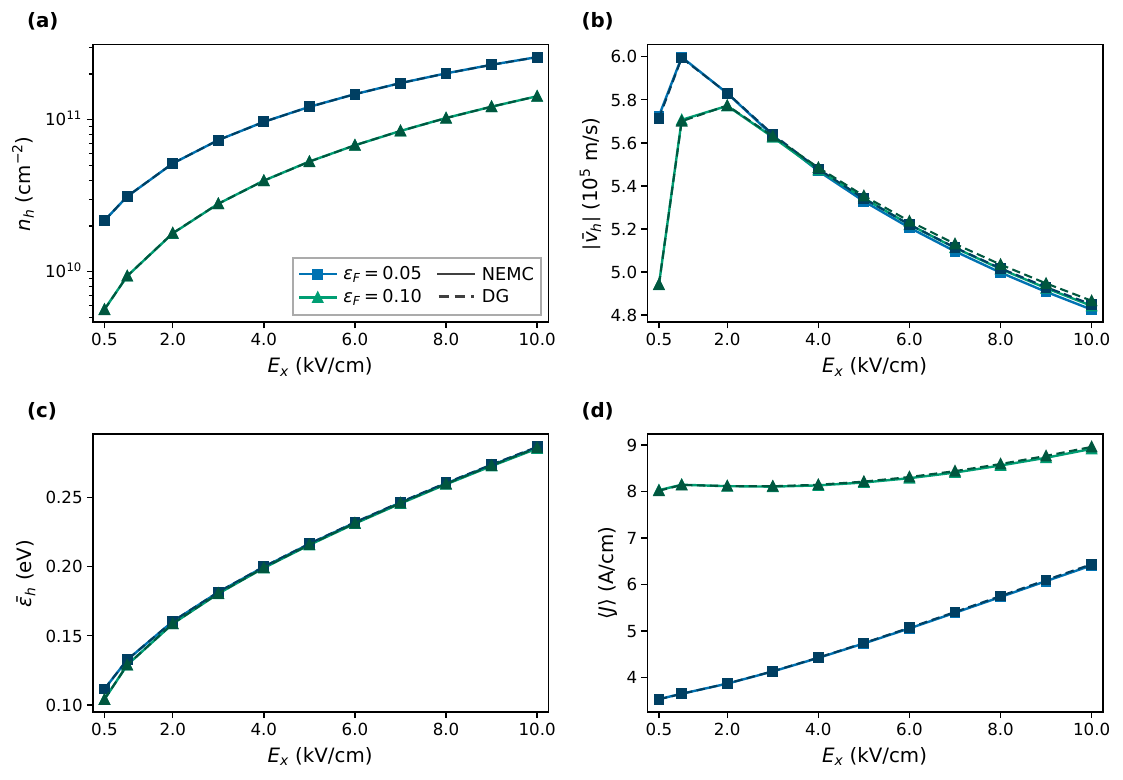}
\caption{Field-dependent NEMC--DG results at
$\varepsilon_F=0.05$ and $0.10$ eV, averaged over
$300$--$400$ ps.
(a) Hole density;
(b) hole drift-velocity magnitude;
(c) mean hole energy;
(d) total current density.}
\label{fig:intermediate-field}
\end{figure}

Additional NEMC--DG comparisons of time-dependent transport observables at charge neutrality and
finite doping are shown in Figs.~\ref{fig:neutral-additional}
and~\ref{fig:additional-transients}, respectively.
The corresponding field-dependent comparisons at
$\varepsilon_F=0.05$ and $0.10$ eV are presented in
Fig.~\ref{fig:intermediate-field}.

The following numerical-sensitivity tests use $200$ ps
simulations, with late-time statistics evaluated over
$160$--$200$ ps, unless stated otherwise. For an observable
$X$, let $X_b$ denote its mean over temporal block
$b=1,\ldots,B$, with $B=4$ nonoverlapping blocks of
$10$ ps. The late-time mean and its block standard error
of the mean (SEM) are
\[
\langle X\rangle=\frac{1}{B}\sum_{b=1}^{B}X_b,
\qquad
\mathrm{SEM}(X)=
\sqrt{\frac{1}{B(B-1)}
\sum_{b=1}^{B}(X_b-\langle X\rangle)^2}.
\]
The relative SEM is
$\mathrm{Rel.\ SEM}(X)=
100\,\mathrm{SEM}(X)/|\langle X\rangle|$
in percent. The relative temporal standard deviation in
Fig.~\ref{fig:nemc-sensitivity}(b) is
$100\,\sigma_t(X)/|\langle X\rangle|$
in percent, where $\sigma_t(X)$ is the standard
deviation of the recorded late-time values of $X$.

The sensitivity to the generation-quadrature order
$q_{\mathrm{gen}}$ was examined for three representative
cases, with results given in Table~\ref{tab:d-gen-quadrature}.
The accepted generation and recombination rates
per unit area and time are
$R_{\mathrm{gen}}=w_p N_{\mathrm{gen}}^{\mathrm{acc}}/\Delta T$
and
$R_{\mathrm{rec}}=w_p N_{\mathrm{rec}}^{\mathrm{acc}}/\Delta T$,
where $N_{\mathrm{gen}}^{\mathrm{acc}}$ and
$N_{\mathrm{rec}}^{\mathrm{acc}}$ are the respective
numbers of accepted events during the observation
interval $\Delta T$.

\begin{table}[pos=htbp]
\centering
\caption{Generation-quadrature sensitivity: late-time
means $\pm$ block SEM over $160$--$200$ ps and total runtimes.}
\label{tab:d-gen-quadrature}
\footnotesize
\renewcommand{\arraystretch}{1.15}
\begin{tabular*}{\textwidth}{@{\extracolsep{\fill}}rrrrrr@{}}
\toprule
$q_{\rm gen}$
& \shortstack{$\langle n_h\rangle$\\
  $(10^9\,\mathrm{cm}^{-2})$}
& \shortstack{$\langle J\rangle$\\
  $(\mathrm{A/cm})$}
& \shortstack{$R_{\rm gen}$\\
  $(10^9\,\mathrm{cm}^{-2}\mathrm{ps}^{-1})$}
& \shortstack{$R_{\rm rec}$\\
  $(10^9\,\mathrm{cm}^{-2}\mathrm{ps}^{-1})$}
& \shortstack{Runtime\\$(\mathrm{h})$} \\
\midrule
\multicolumn{6}{@{}l}{$\varepsilon_F=0$ eV,
                         $E_x=1$ kV/cm} \\
1 & \(108.62\pm 0.011\) & \(2.1267\pm 0.00022\)
  & \(4.4355\pm 0.0022\) & \(4.435\pm 0.0032\) & 5.06 \\
2 & \(108.65\pm 0.02\) & \(2.127\pm 0.00029\)
  & \(4.4375\pm 0.00083\) & \(4.436\pm 0.0025\) & 5.459 \\
3 & \(108.65\pm 0.019\) & \(2.1271\pm 0.0004\)
  & \(4.4365\pm 0.0017\) & \(4.4376\pm 0.0033\) & 5.041 \\
4 & \(108.61\pm 0.013\) & \(2.1265\pm 0.00029\)
  & \(4.4356\pm 0.0011\) & \(4.4338\pm 0.0027\) & 5.289 \\
\addlinespace[2pt]

\multicolumn{6}{@{}l}{$\varepsilon_F=0.05$ eV,
                         $E_x=2$ kV/cm} \\
1 & \(51.616\pm 0.03\) & \(3.8667\pm 0.00068\)
  & \(4.2366\pm 0.0033\) & \(4.2362\pm 0.0053\) & 4.317 \\
2 & \(51.566\pm 0.03\) & \(3.8659\pm 0.00066\)
  & \(4.2389\pm 0.0064\) & \(4.2355\pm 0.0015\) & 4.289 \\
3 & \(51.584\pm 0.017\) & \(3.8662\pm 0.00027\)
  & \(4.2414\pm 0.0051\) & \(4.2424\pm 0.0029\) & 5.098 \\
4 & \(51.616\pm 0.019\) & \(3.8668\pm 0.00049\)
  & \(4.2436\pm 0.0039\) & \(4.2412\pm 0.0045\) & 6.359 \\
\addlinespace[2pt]

\multicolumn{6}{@{}l}{$\varepsilon_F=0.20$ eV,
                         $E_x=2$ kV/cm} \\
1 & \(2.4405\pm 0.014\) & \(21.8415\pm 0.0017\)
  & \(1.3895\pm 0.0038\) & \(1.3897\pm 0.0056\) & 4.538 \\
2 & \(2.4477\pm 0.025\) & \(21.8402\pm 0.0015\)
  & \(1.3823\pm 0.0044\) & \(1.3858\pm 0.011\) & 4.963 \\
3 & \(2.4562\pm 0.013\) & \(21.8410\pm 0.00091\)
  & \(1.3838\pm 0.0045\) & \(1.3829\pm 0.0045\) & 6.19 \\
4 & \(2.4687\pm 0.013\) & \(21.8396\pm 0.0012\)
  & \(1.3995\pm 0.0018\) & \(1.3991\pm 0.0027\) & 7.039 \\
\bottomrule
\end{tabular*}
\end{table}

\begin{figure}[pos=htbp]
\centering
\includegraphics[width=0.75\linewidth]{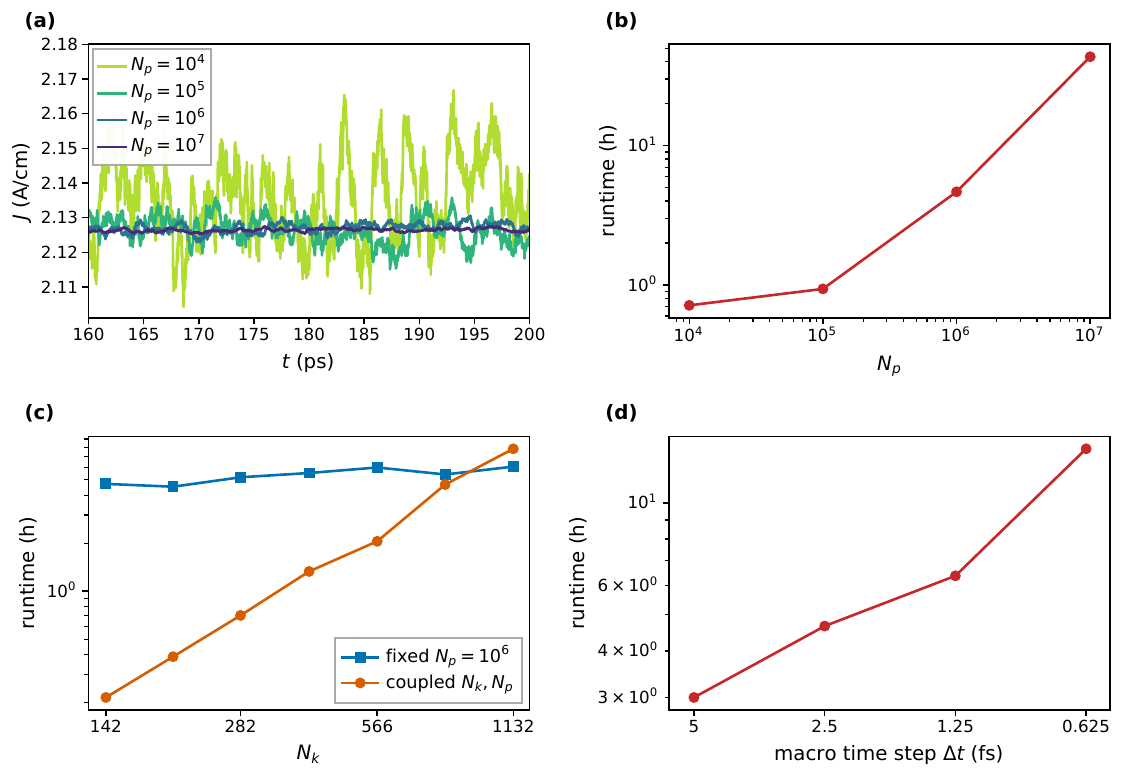}
\caption{Additional NEMC numerical-sensitivity results at
$\varepsilon_F=0$ and $E_x=1$ kV/cm.
(a) Current-density traces over $160$--$200$ ps for
$N_p=10^4$, $10^5$, $10^6$, and $10^7$ at $N_k=800$;
(b) runtime versus $N_p$;
(c) runtime versus $N_k$ for fixed $N_p=10^6$ and
coupled $(N_k,N_p)$ refinement;
(d) runtime versus macro timestep at $N_k=800$
and $N_p=10^6$.}
\label{fig:sensitivity-additional}
\end{figure}

The ensemble-size results at fixed $N_k=800$,
$\varepsilon_F=0$, and $E_x=1$ kV/cm are reported in
Table~\ref{tab:np-sensitivity}.
The corresponding current traces and runtimes are shown
in Fig.~\ref{fig:sensitivity-additional}(a,b).

\begin{table}[pos=htbp]
\centering
\caption{Fixed-grid NEMC ensemble-size study at
$\varepsilon_F=0$, $E_x=1$ kV/cm, and $N_k=800$.
Means $\pm$ block SEM refer to $160$--$200$ ps.}
\label{tab:np-sensitivity}
\footnotesize
\renewcommand{\arraystretch}{1.15}
\begin{tabular*}{\textwidth}{@{\extracolsep{\fill}}rrrrrr@{}}
\toprule
$N_p$
& \shortstack{$\langle n_e\rangle$\\
  $(10^{11}\,\mathrm{cm}^{-2})$}
& \shortstack{$\langle J\rangle$\\
  $(\mathrm{A/cm})$}
& \shortstack{$\langle\bar{\varepsilon}_e\rangle$\\
  $(\mathrm{eV})$}
& \shortstack{Rel. SEM($J$)\\$(\%)$}
& \shortstack{Runtime\\$(\mathrm{h})$} \\
\midrule
$10^4$
& $1.0919\pm0.0008$
& $2.1358\pm0.0015$
& $0.13518\pm5\times10^{-5}$
& 0.0714
& 0.7144 \\

$10^5$
& $1.0859\pm0.00049$
& $2.1264\pm0.00092$
& $0.1352\pm1.4\times10^{-5}$
& 0.0431
& 0.9376 \\

$10^6$
& $1.0865\pm0.0002$
& $2.1270\pm0.00029$
& $0.13516\pm7.3\times10^{-6}$
& 0.0135
& 4.664 \\

$10^7$
& $1.0860\pm4.5\times10^{-5}$
& $2.1264\pm0.00011$
& $0.13516\pm2.6\times10^{-6}$
& 0.00534
& 43.43 \\
\bottomrule
\end{tabular*}
\end{table}

Additional NEMC occupation-grid and macro-timestep
refinement results for six numerical-test cases are
collected in Table~\ref{tab:d-nemc}. The percentage changes
are measured between successive discretizations relative
to the refined values, rather than to an assumed exact
solution. For drift velocity and mean energy, the larger
of the electron and hole percentage changes is reported.
For the current, the dimensional difference between
successive refinements is
$\delta J=\langle J\rangle_{\mathrm{refined}}
-\langle J\rangle_{\mathrm{coarse}}$.
The corresponding relative change is
$\Delta J=100|\delta J|/
|\langle J\rangle_{\mathrm{refined}}|$
in percent, while the SEM-normalized difference is
$|\delta J|/\mathrm{SEM}_{\mathrm{comb}}$.
The combined block SEM is
$\mathrm{SEM}_{\mathrm{comb}} =
(\mathrm{SEM}_{J,\mathrm{coarse}}^2+
\mathrm{SEM}_{J,\mathrm{refined}}^2)^{1/2}$;
it normalizes the observed current change and is not an
independent-run significance test. The runtimes for the
ensemble-size, grid-refinement, and macro-timestep tests
are compared in Fig.~\ref{fig:sensitivity-additional}.

\begin{table}[pos=htbp]
\centering
\caption{NEMC refinement sensitivity over $160$--$200$ ps.
The current shift is normalized by the combined block SEM;
runtime ratios are refined/coarser.}
\label{tab:d-nemc}
\scriptsize
\setlength{\tabcolsep}{2pt}
\renewcommand{\arraystretch}{1.12}
\begin{tabular*}{\textwidth}{@{\extracolsep{\fill}}rrrrrrrrr@{}}
\toprule
\shortstack{$\varepsilon_F$\\(eV)}
& \shortstack{$E_x$\\(kV/cm)}
& \shortstack{$\Delta n_e$\\(\%)}
& \shortstack{$\Delta n_h$\\(\%)}
& \shortstack{$\Delta v$\\(\%)}
& \shortstack{$\Delta\bar\varepsilon$\\(\%)}
& \shortstack{$\Delta J$\\(\%)}
& \shortstack{$|\delta J|/$\\$\mathrm{SEM}_{\rm comb}$}
& \shortstack{$t_{\rm new}/$\\$t_{\rm old}$} \\
\midrule

\multicolumn{9}{@{}l}{Fixed $N_p=10^6$:
$N_k:800\rightarrow1132$} \\
0    & 0.5 & 0.0657  & 0.0657 & 0.0108   & 0.00972
     & 0.0589 & 1.13  & 1.13 \\
0    & 1   & 0.00368 & 0.00368 & 0.0106   & 0.00369
     & 0.00893 & 0.38 & 1.12 \\
0    & 2   & 0.0279  & 0.0279 & 0.013    & 0.00509
     & 0.0359 & 1.30  & 1.09 \\
0.05 & 0.5 & 0.0121  & 0.189  & 0.00981  & 0.0624
     & 0.031  & 1.40  & 1.11 \\
0.05 & 1   & 0.0122  & 0.136  & 0.0128   & 0.0508
     & 0.0178 & 0.666 & 0.851 \\
0.05 & 2   & 0.0179  & 0.128  & 0.000714 & 0.0136
     & 0.031  & 1.21  & 0.789 \\

\addlinespace[3pt]
\multicolumn{9}{@{}l}{Coupled $(N_k,N_p):
(800,10^6)\rightarrow(1132,2\times10^6)$} \\
0    & 0.5 & 0.122   & 0.122  & 0.0221  & 0.00646
     & 0.110   & 2.76  & 2.22 \\
0    & 1   & 0.0157  & 0.0157 & 0.00572 & 0.00116
     & 0.0111  & 0.493 & 1.68 \\
0    & 2   & 0.0220  & 0.0220 & 0.00881 & 0.000235
     & 0.0203  & 0.675 & 2.08 \\
0.05 & 0.5 & 0.00380 & 0.0594 & 0.0937  & 0.00914
     & 0.00486 & 0.386 & 2.13 \\
0.05 & 1   & 0.00804 & 0.0897 & 0.0192  & 0.0330
     & 0.0319  & 1.36  & 1.56 \\
0.05 & 2   & 0.0175  & 0.126  & 0.00367 & 0.00472
     & 0.0285  & 1.14  & 2.10 \\

\addlinespace[3pt]
\multicolumn{9}{@{}l}{Macro step:
$\Delta t:1.25\rightarrow0.625$ fs} \\
0    & 0.5 & 0.0584  & 0.0584 & 0.0369 & 0.0288
     & 0.0357 & 0.946 & 2.61 \\
0    & 1   & 0.0394  & 0.0394 & 0.0210 & 0.0115
     & 0.0585 & 3.25  & 2.20 \\
0    & 2   & 0.0433  & 0.0433 & 0.0437 & 0.0257
     & 0.0854 & 3.47  & 2.05 \\
0.05 & 0.5 & 0.00786 & 0.123  & 0.0106 & 0.0188
     & 0.0192 & 0.796 & 2.26 \\
0.05 & 1   & 0.0191  & 0.213  & 0.0275 & 0.00787
     & 0.0619 & 3.22  & 2.17 \\
0.05 & 2   & 0.00658 & 0.0471 & 0.0429 & 0.0249
     & 0.0296 & 1.89  & 2.24 \\
\bottomrule
\end{tabular*}
\end{table}

\begin{figure}[pos=htbp]
\centering
\includegraphics[width=0.75\linewidth]{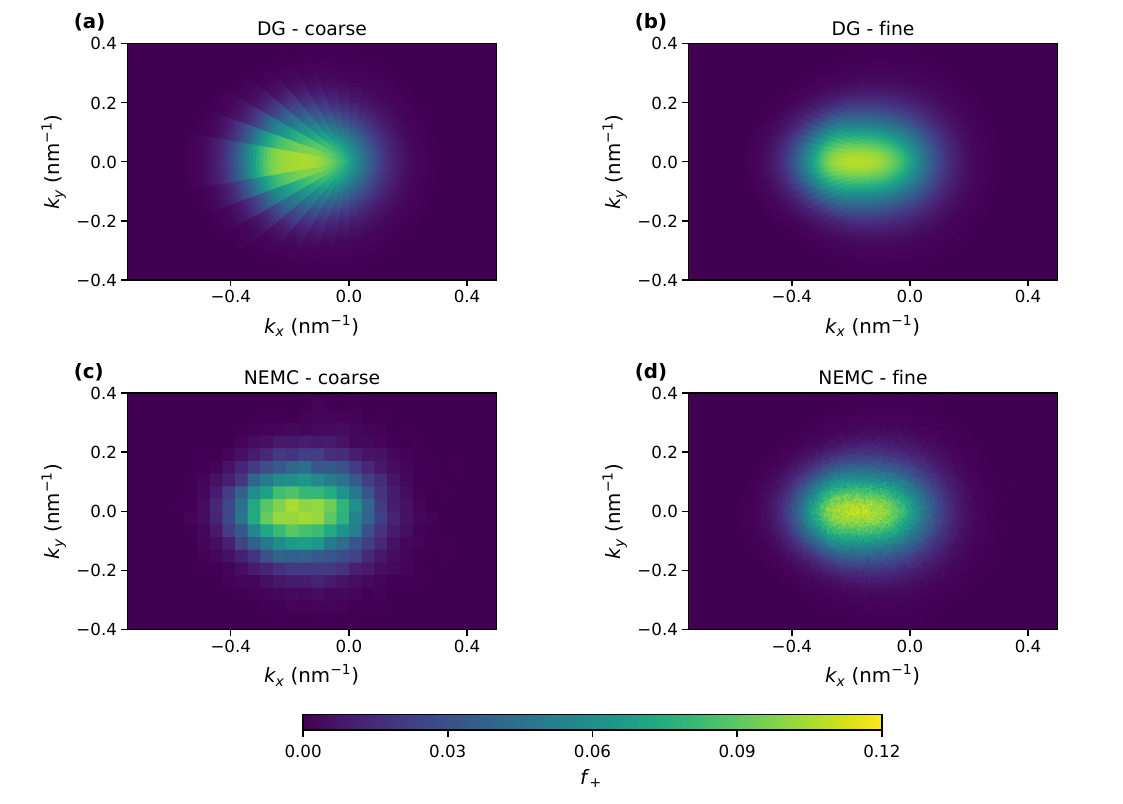}
\caption{Electron occupation $f_+(k_x,k_y)$ at
$t=200$ ps for $\varepsilon_F=0$ and $E_x=1$ kV/cm.
(a,b) DG with $N_E\times N_\theta=250\times36$
and $707\times102$;
(c,d) NEMC with $(N_k,N_p)=(142,31250)$
and $(1132,2\times10^6)$.
All panels use the same color scale.}
\label{fig:occupation-maps}
\end{figure}

Figure~\ref{fig:occupation-maps} compares the final
electron occupation distributions for selected coarse and
refined DG and NEMC discretizations.

\begin{figure}[pos=htbp]
\centering
\includegraphics[width=0.75\linewidth]{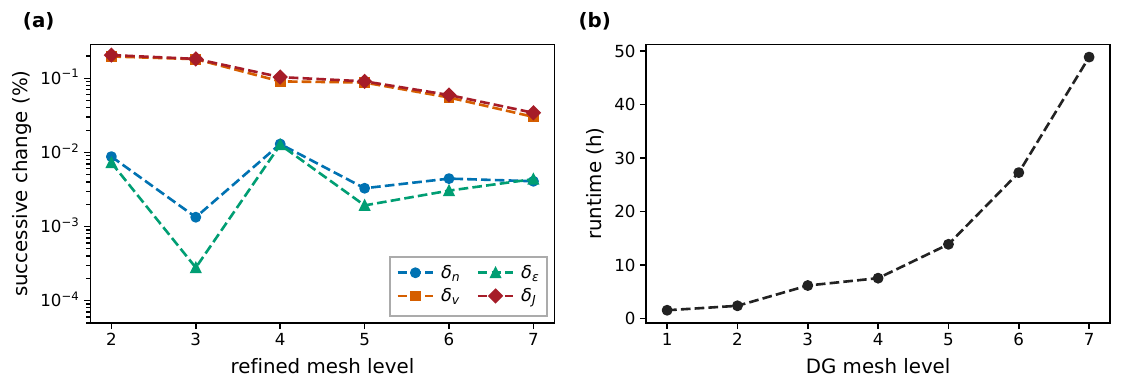}
\caption{DG mesh refinement at $\varepsilon_F=0$ and
$E_x=1$ kV/cm. The mesh levels $L=1,\ldots,7$
correspond to $N_E\times N_\theta=
250\times36$, $297\times42$, $354\times50$,
$420\times60$, $500\times72$, $595\times86$,
and $707\times102$, respectively.
(a) Relative changes between levels $L-1$ and $L$;
(b) runtime.}
\label{fig:dg-refinement}
\end{figure}

The DG mesh-refinement results are presented in
Fig.~\ref{fig:dg-refinement} and
Table~\ref{tab:d-dg}.

\begin{table}[pos=htbp]
\centering
\caption{DG refinement sensitivity. Percentage changes are
relative to the refined values; runtime ratios are
refined/coarser. For drift velocity and mean energy,
the larger electron/hole change is listed.}
\label{tab:d-dg}
\scriptsize
\renewcommand{\arraystretch}{1.15}
\begin{tabular*}{\textwidth}{@{\extracolsep{\fill}}rrrrrrrrr@{}}
\toprule
\shortstack{$\varepsilon_F$\\(eV)}
& \shortstack{$E_x$\\(kV/cm)}
& \shortstack{$\Delta n_e$\\(\%)}
& \shortstack{$\Delta n_h$\\(\%)}
& \shortstack{$\Delta v$\\(\%)}
& \shortstack{$\Delta\bar{\varepsilon}$\\(\%)}
& \shortstack{$\Delta J$\\(\%)}
& \shortstack{$t_{\rm new}/$\\$t_{\rm old}$}
& \shortstack{$\langle\Delta t\rangle$\\(fs)} \\
\midrule

\multicolumn{9}{@{}l}{Coupled DG mesh:
$595\times86\rightarrow707\times102$} \\
0    & 0.5 & 0.00156   & 0.00156 & 0.0373 & 0.00351
     & 0.0389 & 2.40 & $0.833\to0.500$ \\
0    & 1   & 0.00408   & 0.00408 & 0.0302 & 0.00437
     & 0.0343 & 1.79 & $0.417\to0.278$ \\
0    & 2   & 0.00745   & 0.00745 & 0.0251 & 0.00518
     & 0.0325 & 1.67 & $0.208\to0.147$ \\
0.05 & 0.5 & 0.0000838 & 0.00131 & 0.0370 & 0.00341
     & 0.0364 & 2.46 & $0.833\to0.500$ \\
0.05 & 1   & 0.000662  & 0.00738 & 0.0309 & 0.00469
     & 0.0307 & 1.78 & $0.417\to0.278$ \\
0.05 & 2   & 0.00188   & 0.0134  & 0.0255 & 0.00541
     & 0.0281 & 1.68 & $0.208\to0.147$ \\

\addlinespace[3pt]
\multicolumn{9}{@{}l}{Radial refinement:
$500\times72\rightarrow750\times72$} \\
0 & 1 & 0.00260 & 0.00260 & 0.0300 & 0.00359
  & 0.0274 & 2.68 & $0.500\to0.357$ \\

\addlinespace[3pt]
\multicolumn{9}{@{}l}{Angular refinement:
$500\times72\rightarrow500\times108$} \\
0 & 1 & 0.0128 & 0.0128 & 0.0625 & 0.0124
  & 0.0753 & 2.34 & $0.500\to0.357$ \\
\bottomrule
\end{tabular*}
\end{table}

The remaining initialization, finite-domain, and
numerical-consistency checks are summarized in
Table~\ref{tab:d-checks}.
The maximum deviation from the charge invariant
over a simulation of duration $T$ is
\[
\Delta_{\mathrm{charge}}=
\max_{0\leq t\leq T}
\left|
[n_e(t)-n_h(t)]
-[n_e(0)-n_h(0)]
\right|.
\]
The domain-rejection fraction is
$r_{\mathrm{dom}}=
N_{\mathrm{dom}}^{\mathrm{rej}}/N_{\mathrm{prop}}$,
where $N_{\mathrm{dom}}^{\mathrm{rej}}$ is the number
of proposals rejected because their targets lie outside
$\Omega_k$, and $N_{\mathrm{prop}}$ is the total
number of collision-event proposals.

\begin{table}[pos=htbp]
\centering
\caption{Initialization, finite-domain, and
numerical-consistency diagnostics.}
\label{tab:d-checks}
\footnotesize
\setlength{\tabcolsep}{4pt}
\renewcommand{\arraystretch}{1.12}

\begin{tabularx}{\textwidth}{@{}
>{\raggedright\arraybackslash}p{0.27\textwidth}
>{\raggedright\arraybackslash}X
>{\raggedright\arraybackslash}p{0.32\textwidth}
@{}}
\toprule
Condition & Diagnostic & Result \\
\midrule

\multicolumn{3}{@{}l}{\emph{Initialization:
$q_{\rm init}=4$ versus $16$}} \\

Four initial Fermi energies
& Maximum relative density change
& $3.308\times10^{-5}\%$
  ($n_h$, $\varepsilon_F=0.20$ eV) \\

Four initial Fermi energies
& Maximum cell-occupation difference
& $3.468\times10^{-6}$
  ($f_+$, $\varepsilon_F=0$) \\

$\varepsilon_F=0$ eV
& Initial electron/hole counts
& $1{,}000{,}058/1{,}000{,}058$ \\

$\varepsilon_F=0.05$ eV
& Initial electron/hole counts
& $1{,}920{,}432/79{,}444$ \\

$\varepsilon_F=0.10$ eV
& Initial electron/hole counts
& $1{,}995{,}923/4{,}557$ \\

$\varepsilon_F=0.20$ eV
& Initial electron/hole counts
& $1{,}999{,}145/28$ \\

\addlinespace[3pt]
\multicolumn{3}{@{}l}{\emph{Finite-domain tests:
$\varepsilon_F=0$, $E_x=1$ kV/cm}} \\

NEMC: $\varepsilon_{\max}:2\to3$ eV;
$N_k:800\to1200$
& Relative changes in $(n_e,n_h,J)$;
runtime ratio
& $(0.02224,\ 0.02224,\ 0.007857)\%$;
$t_3/t_2=1.328$ \\

DG: $\varepsilon_{\max}:2\to3$ eV;
$N_E:500\to750$
& Relative changes in $(n_e,n_h,J)$;
runtime ratio
& $(1.265\times10^{-13},\
1.405\times10^{-13},\ 0)\%$;
$t_3/t_2=1.724$ \\

Production snapshots,
$200$--$400$ ps
& Maximum fraction of electrons
outside the DG disk
& $0$ \\

\addlinespace[3pt]
\multicolumn{3}{@{}l}{\emph{Discrete consistency:
200 ps simulations}} \\

NEMC: $\varepsilon_F=0$,
$E_x=1$ kV/cm,
$q_{\rm gen}=1$
& Outer conduction/hole occupations
& $0/0$ \\

Same NEMC case
& Minimum/maximum recorded
final occupation
& $0/1$ \\

NEMC: $\varepsilon_F=0.20$ eV,
$E_x=2$ kV/cm,
$q_{\rm gen}=4$
& Domain-rejection fraction
& $6.4982\times10^{-9}$ \\

NEMC: $\varepsilon_F=0.20$ eV,
$E_x=2$ kV/cm,
$q_{\rm gen}=1$
& Maximum charge-invariant deviation
(cm$^{-2}$)
& $4.8828\times10^{-4}$ \\

DG: $\varepsilon_F=0$,
$E_x=1$ kV/cm,
$500\times72$ grid
& Maximum charge-invariant deviation
(cm$^{-2}$)
& $4.25\times10^{-4}$ \\

\bottomrule
\end{tabularx}
\end{table}

\FloatBarrier

\section*{Declaration of generative AI and
AI-assisted technologies in the manuscript
preparation process}

During the preparation of this work, the authors
used ChatGPT (OpenAI) to assist with improving
the language, clarity, and organization of the
manuscript. The authors reviewed and edited
the resulting content and take full responsibility
for the content of the publication.

\bibliographystyle{unsrtnat}
\bibliography{Zalinyan_Nastasi_refs}
\end{document}